\documentclass{article}
\usepackage{arxiv}          
\usepackage[onecolumn,letterpaper]{arxivcompat}

\usepackage{lipsum}
\usepackage{booktabs}
\usepackage{caption}
\usepackage{subcaption}
\usepackage{makecell}

\usetikzlibrary{arrows.meta, shapes.geometric}
\newif\ifreviewcomments
\reviewcommentsfalse

\title{Varda-single-1.0: deterministic data-driven weather forecasting at 1\,km resolution over Switzerland's complex topography}
\author{\authorblock{%
  \authorname{Alberto Pennino}\affilmark{1,\dag},
  \authorname{Francesco Zanetta}\affilmark{1,\dag},
  \authorname{Michele Cattaneo}\affilmark{2,1},
  \authorname{Claire Merker}\affilmark{1},
  \authorname{Radi Radev}\affilmark{1},
  \authorname{Jonas Bhend}\affilmark{1},
  \authorname{Louis Frey}\affilmark{1},
  \authorname{Hugues de Laroussilhe}\affilmark{1},
  \authorname{Ophélia Miralles}\affilmark{1},
  \authorname{Carlos Osuna}\affilmark{1},
  \authorname{Daniele Nerini}\affilmark{1},
  \authorname{Andreas Pauling}\affilmark{1},
  \authorname{Daniel Hupp}\affilmark{1},
  \authorname{Ulrich Hamann}\affilmark{1},
  \authorname{Mary McGlohon}\affilmark{1},
  \authorname{Marti Bosch}\affilmark{3,1},
  \authorname{Luca Lanzilao}\affilmark{3,1},
  \authorname{Marco Arpagaus}\affilmark{1},
  \authorname{Lukas Jansing}\affilmark{1},
  \authorname{Daniel Leuenberger}\affilmark{1},
  \authorname{Mark A. Liniger}\affilmark{1},
  \authorname{Katrin Ehlert}\affilmark{1},
  \authorname{Matthew Chantry}\affilmark{4},
  \authorname{Håvard Homleid Haugen}\affilmark{5},
  \authorname{Gert Mertes}\affilmark{4},
  \authorname{Ana Prieto Nemesio}\affilmark{4},
  \authorname{Mario Santa Cruz}\affilmark{4},
  \authorname{Jasper Wijnands}\affilmark{6},
  \authorname{Gabriel Moldovan}\affilmark{4},
  \authorname{Harrison Cook}\affilmark{4},
  and \authorname{Oliver Fuhrer}\affilmark{1,7}\\[0.6em]
  \affilline{1}{Federal Office of Meteorology and Climatology MeteoSwiss}\\
  \affilline{2}{Swiss Data Science Center (SDSC), ETH Zürich}\\
  \affilline{3}{Center for Climate Systems Modeling (C2SM), ETH Zürich}\\
  \affilline{4}{European Centre for Medium-Range Weather Forecasts (ECMWF)}\\
  \affilline{5}{Norwegian Meteorological Institute (MET Norway)}\\
  \affilline{6}{Royal Netherlands Meteorological Institute (KNMI)}\\
  \affilline{7}{Institute for Atmospheric and Climate Science (IAC), ETH Zürich}\\[0.4em]
  \authornote{\affilmark{\dag}Correspondence: \mbox{Alberto Pennino (\texttt{alberto.pennino@meteoswiss.ch})} and \mbox{Francesco Zanetta (\texttt{francesco.zanetta@meteoswiss.ch})}}\\
  \authornote{These authors contributed equally: Alberto Pennino, Francesco Zanetta}
}}
\date{\today}
\renewcommand{\shorttitle}{Varda-single}
\renewcommand{\headeright}{A Preprint}
\renewcommand{\undertitle}{A Preprint}

\begin{document}
\maketitle

\begin{abstract}

We present Varda-single-1.0, a medium-range data-driven weather prediction system built for the Alpine domain. It provides hourly deterministic regional forecasts on a mesh of 1\,\unit{km} resolution and global forecasts on a 31\,\unit{km} mesh. The system comprises two independently trained stretched-grid Graph Transformer models with encoder--processor--decoder architecture, developed in the Anemoi framework: a 6-hourly autoregressive forecaster and a temporal downscaler reconstructing hourly forecasts between the forecaster's steps. Its training curriculum includes pre-training on ERA5 reanalysis data, followed by training on a 20-year kilometre-scale regional reanalysis, and finally fine-tuning on operational kilometre-scale analyses.
Verified over one year against operational analyses and surface station observations, Varda-single is competitive with or improves on MeteoSwiss' operational numerical weather prediction baselines for most headline scores and variables. It broadly matches the skill of the high-resolution 1\,\unit{km} ICON-CH1-EPS control at  lead times up to +33\,\unit{h} and generally outperforms the 2\,\unit{km} ICON-CH2-EPS control at lead times up to +120\,\unit{h}. Despite its competitive aggregate scores, Varda-single underestimates some local wind maxima and produces overly smooth convective precipitation fields, consistent with the smoothing associated with squared-error training. In order to gain insight into the model's behaviour, we investigate three case studies in addition to the aggregated headline scores, and find particular weaknesses in Varda-single’s representation of local winds over complex terrain.
The introduction of Varda-single represents an important step in the development of high resolution ML forecasting over complex terrain, in complementing the operational regional numerical weather prediction models of MeteoSwiss with data-driven models and in providing a pretrained model for researchers and user-specific applications.

\end{abstract}

\section{Introduction}

\begin{figure*}[!htbp]
    \centering
    \includegraphics[scale=0.9, trim = 0 0 0 20, clip]{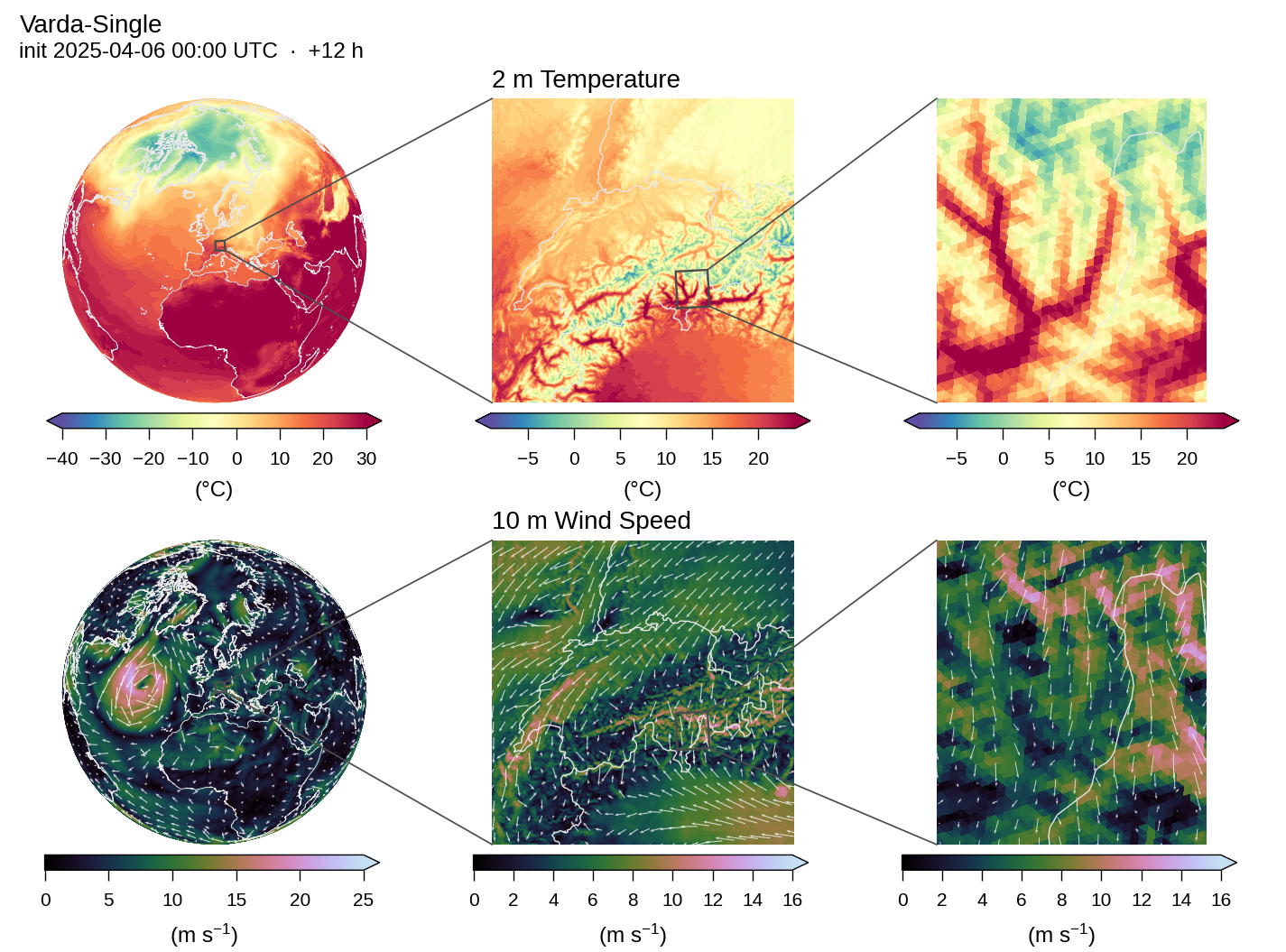}
    \caption{Example of a Varda-single forecast initialised at 00~UTC on 6 April 2025, showing 2\,\unit{m} temperature and 10\,\unit{m} wind speed at a lead time of +12\,\unit{h}.}
    \label{fig:teaser}
\end{figure*}

Data-driven weather forecasting has advanced rapidly over the past few years. Global deterministic prediction was targeted first, with several models now matching or exceeding the skill of leading numerical weather prediction (NWP) systems in the medium range, at a small fraction of the inference cost \citep{keisler2022, bi2023n, lam2023sa, chen2023ncas, lang2024, bodnar2024, bouallegue2024bams, rasp2024}. Research work has since extended in two directions: probabilistic and generative formulations \citep{price2024, lang2024, alet2025, bonev2025} and regional forecasting at higher resolution \citep{oskarsson2023a, adamov2025, nipen2025aies, wijnands2025, miralles2026-aies, smith2026bridgingshortmediumrangeweather, miralles2026}.

Regional forecasting in particular is where national meteorological services (NMS) focus most of their efforts: the forecasts they are responsible for depend on processes acting on scales of a few kilometres - such as convection, orographic and valley winds or fog, to name a few - that are not resolved at the spatial resolutions typical for global models. Therefore, NMS operate high-resolution limited-area NWP systems, often as ensembles and with frequent updates and hourly or sub-hourly output, which require substantial computing resources \citep{lapillonne2026gmd}. The development of data-driven alternatives for these systems has therefore naturally become a focus of several NMS.

Currently, two main approaches have been proposed for building data-driven models at high spatial resolution. The limited-area modelling (LAM) approach, as proposed by \citet{oskarsson2023a}, defines a network covering only the regional domain, with lateral boundary conditions provided by an external model. An alternative is the stretched-grid approach, as used in the Bris model \citep{nipen2025aies}, which uses a single model over a global domain with a locally refined grid and adopts a curriculum learning strategy spanning multiple grids and graph structures. Both lines of work build on the encoder--processor--decoder graph neural network (GNN) architecture of \citet{keisler2022} and \citet{lam2023sa}, which is flexible enough to handle irregular and mixed-resolution grids. An extensive comparison of the two approaches is given by \citet{wijnands2025}.

Beyond the structural grid differences discussed above, both approaches share a common challenge: they operate at 6-hourly time steps. At this resolution, many extremes occurring between output steps are missed, the diurnal cycle and its related processes over complex orography are not well represented, and the life cycles of processes such as convection are unresolved. Numerous downstream users depend on finer temporal resolution. Examples include aviation, where short-term turbulence, wind shear, and icing forecasts are safety-critical; the energy sector, where solar and wind power forecasting and grid balancing require accurate information on rapid temporal variability; and hydrology, where flash-flood warnings require rainfall intensity to be resolved  at hourly or finer temporal resolution. Recent work addresses this gap either with a separate model that fills in the intermediate states \citep{leinonen2024, zhong2024, ingstad2026}, or by coupling the learned model to a physical solver at a shorter time step \citep{xu2024}.

Much of this work has been consolidated in the open-source software framework Anemoi \citep{lang2024}, developed by the European Centre for Medium-Range Weather Forecasts (ECMWF) together with a growing community of European NMS \citep[see e.g.][]{nipen2025aies, lang2024a, ingstad2026, goecke2026aiconoperationalglobalmachine, miralles2026-aies} and other institutions \citep[][]{bano-medina2025, smith2026bridgingshortmediumrangeweather}. Anemoi covers the full model lifecycle, from dataset creation and model development to training and inference. By providing a common framework for these stages, it lowers the engineering effort required to develop operational data-driven forecasting systems, allowing users to focus instead on adapting models to their specific forecasting requirements.

MeteoSwiss, Switzerland's NMS, operates ICON-CH1-EPS and ICON-CH2-EPS, two regional ensemble configurations of the ICON model \citep{zangl2015} running at approximately 1\,\unit{km} and 2\,\unit{km} horizontal resolution. They are initialised every 3 and 6 hours, with forecast horizons of +33\,\unit{h} and +120\,\unit{h}, respectively. These systems are the basis for warnings over the Swiss Alps, a demanding forecasting environment where steep orography drives rapid mesoscale circulations, valley-channelled flows, and convective triggering on scales of a few kilometres. MeteoSwiss' deterministic NWP forecasts and derived products are well-established, of high quality and widely used. This, in combination with the overall smaller size of the problem as compared to a probabilistic model, motivated our choice of developing a deterministic system first, paving the way for future probabilistic extensions.

We present Varda-single-1.0, a deterministic forecasting system developed with Anemoi, producing forecasts at 1\,\unit{km} spatial resolution and at 1-hour temporal resolution over the Alpine domain, as illustrated in \autoref{fig:teaser}. The figure shows both the global context and the high-resolution local detail of a Varda-single forecast. Varda-single consists of two components: an autoregressive forecasting model, trained to predict 6\,\unit{h} increments, and a temporal downscaling model, trained to interpolate hourly values given bounding states of 6\,\unit{h} windows (\autoref{fig:varda}). Both components follow the encoder--processor--decoder stretched-grid architecture, operating on a global 31\,\unit{km} mesh refined to 1\,\unit{km} over Switzerland and the surrounding region. The preference for a stretched-grid approach over LAM was mainly motivated by technical aspects, as the former does not require lateral boundary conditions, without externally supplied forecasts, to operate \citep{wijnands2025}. The forecasting model is trained in several stages, starting as a global model before the regional high-resolution grid is introduced, and ending with a fine-tuning stage on operational regional analyses. 

\begin{table}[htbp]
\centering
\caption{Summary of features included in Varda-single. Variables used by the 6-hourly forecaster are marked with "F", while 
variables used by the temporal downscaler are marked with "D".}
\label{tab:variables}
\setlength{\tabcolsep}{4.5pt}
\begin{tabular}{>{\ttfamily\scriptsize}llll}

\toprule
 & Description & Type  & Model \\
\midrule
\multicolumn{4}{l}{\textbf{Single-level features}} \\
\midrule
2t              & 2\,\unit{m} air temperature          & Prog. & F,D\\
2d              & 2\,\unit{m} dew point temperature    & Prog. & F,D\\
10u             & 10\,\unit{m} eastward wind     & Prog. & F,D\\
10v             & 10\,\unit{m} northward wind     & Prog. & F,D\\
lsm             & Land-sea mask                    & Forc.$^{\star}$ & F,D\\
msl             & Mean sea-level pressure          & Prog. & F,D\\
sp              & Surface pressure                 & Prog. & F,D\\
tp              & Total precipitation (6\,\unit{h}) & Diag. & F\\
tp              & Total precipitation (1\,\unit{h}) & Diag. & D\\
z               & Surface orography                & Forc.$^{\star}$ & F,D\\
sdor            & Subgrid orography std.\ dev.     & Forc.$^{\star}$ & F\\
slor            & Orographic slope                 & Forc.$^{\star}$ & F\\
insolation      & Solar insolation                 & Forc.$^{\dagger}$ & F,D\\
sin\_latitude   & Sine of latitude                 & Forc.$^{\dagger\star}$ & F,D\\
cos\_latitude   & Cosine of latitude               & Forc.$^{\dagger\star}$ & F,D\\
sin\_longitude  & Sine of longitude                & Forc.$^{\dagger\star}$ & F,D\\
cos\_longitude  & Cosine of longitude              & Forc.$^{\dagger\star}$ & F,D\\
cos\_julian\_day & Cosine of Julian day            & Forc.$^{\dagger}$ & F,D\\
sin\_julian\_day & Sine of Julian day              & Forc.$^{\dagger}$ & F,D\\
sin\_local\_time & Sine of local time              & Forc.$^{\dagger}$ & F,D\\
cos\_local\_time & Cosine of local time            & Forc.$^{\dagger}$ & F,D \\
\midrule
\multicolumn{4}{l}{\normalfont\textbf{Pressure-level features on levels 50, 100, 150, 200, 250,}} \\
\multicolumn{3}{l}{\normalfont\textbf{300, 400, 500, 700, 850, 925, and 1000\,\boldmath\unit{hPa}}} \\
\midrule
t\_p & Air temperature                &  Prog. & F,D\\
q\_p & Specific humidity              &  Prog. & F,D\\
u\_p & Eastward wind            &  Prog. & F,D\\
v\_p & Northward wind            &  Prog. & F,D\\
z\_p & Geopotential                   &  Prog. & F,D\\
\bottomrule
\multicolumn{4}{l}{\footnotesize Forc.$^{\dagger}$ - computed analytically;\quad Forc.$^{\star}$ - constant in time}\\
\end{tabular}
\end{table}

\section{Datasets}\label{sec:datasets}

Our training pipeline relies on two global and two regional datasets, paired by training stage: reanalysis products for pre-training and training, and operational analyses for final fine-tuning. The latter match the initial conditions encountered during operations.

Pre-training is performed on ERA5 \citep{hersbach2020qjrms}, used on its native N320 reduced Gaussian grid at approximately 31\,\unit{km} resolution. ERA5 additionally drives the regional downscaling described below at its lateral boundaries. For fine-tuning we use the ECMWF IFS operational analyses, interpolated to the same N320 grid.

\begin{figure*}[!t]
\centering
    \includegraphics[width=0.85\linewidth]{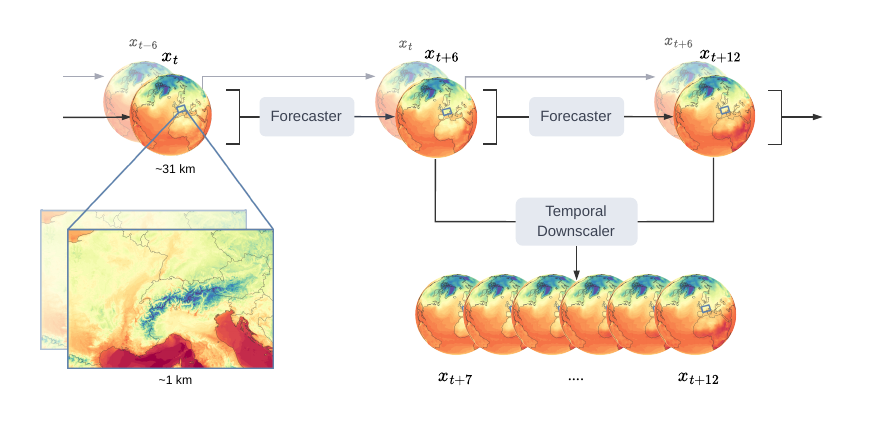}
    \caption{The Varda-single system. The \textit{forecaster} model takes in the current and previous atmospheric states $x_{t,t-6}$ and propagates them forward in time by 6 hours to produce $x_{t+6}$. The \textit{temporal downscaler} model takes in inputs $x_{t}$ and $x_{t+6}$ and produces hourly outputs from $x_{t+1}$ to $x_{t+5}$, also including $x_{t+6}$ for hourly accumulated variables, in our case total precipitation.} 
    \label{fig:varda}
\end{figure*}

Both regional datasets are produced with the ICON model on the MeteoSwiss forecasting domain (whose extent can be seen in \autoref{fig:varda}), sharing the configuration of the operational ICON-CH1-EPS featuring an icosahedral grid of approximately 1\,\unit{km} resolution with terrain-following vertical coordinates. The resulting grid comprises 1,147,980 nodes in the horizontal. 
The first regional dataset, central to this work, is the ICON Reanalysis-Light-CH1 \citep[REA-L-CH1,][]{meteoschweiz2025}, a deterministic dynamical downscaling of ERA5 in which radar precipitation observations are assimilated through latent heat nudging \citep{stephan2008}. It covers January 2005 to March 2025 at 1\,\unit{h} granularity, resulting in a temporal dimension of 177,468 time steps.
The second is the set of operational analyses from the Kilometre-scale Ensemble Data Assimilation \citep[KENDA,][]{schraff2016qjrms} system running at MeteoSwiss (KENDA-CH1), used alongside the IFS analyses in the fine-tuning stage. KENDA-CH1 uses an Ensemble Kalman Filter and assimilates conventional observations (surface stations, radiosondes, aircraft and wind profilers), as well as radar-derived precipitation via latent heat nudging, on a 1-hour cycle. The resulting hourly analyses provide the initial conditions of the operational ICON-CH1-EPS forecasts.

\subsection{Forecasting variables}
At each training stage when the stretched-grid approach is used, we combine a global and a regional dataset, thus requiring a common set of variables between the two. We therefore train the models on the variables available in all datasets (\autoref{tab:variables}), at all stages.
Prognostic variables are atmospheric state variables serving as both model inputs and outputs, diagnostic variables describe the state but enter only as targets, and forcings influence the state without being predicted, being either constant in time or computed analytically at any point in space and time.
Two processing steps are specific to the ICON-based datasets. Since ICON-CH1-EPS does not use pressure coordinates natively, fields are vertically interpolated from model to pressure levels. Where the complex Alpine orography places a pressure level below the surface, values are extrapolated.

\section{Varda-single}\label{sec:varda}

\subsection{System overview}

Varda-single consists of two independently trained graph neural networks: a \emph{6-hourly autoregressive forecaster} that predicts the atmospheric state at 6-hour steps, and a \emph{temporal downscaler} that reconstructs the five intermediate hourly states between any two consecutive 6-hour forecasts, plus a sixth step to account for diagnostic accumulated values at the end of the period (see \autoref{fig:varda}).
The two-stage design is motivated by the trade-off between temporal resolution and error accumulation: an autoregressive model operating at 1-hour steps accumulates errors at every rollout step over a 120-hour forecast, whereas a 6-hourly forecaster requires one-sixth as many sequential rollout steps and the temporal downscaler, conditioned on both endpoints, does not propagate errors forward.
This approach follows the two-stage setup of \citet{zhong2024} and \citet{ingstad2026}, adapting it to the regional operational setting of MeteoSwiss.

\subsection{Graph construction}
\label{sec:graph}

Both the forecaster and the temporal downscaler are Graph Transformer models operating on a graph whose \emph{data nodes} coincide with the training grid points and whose \emph{hidden nodes} form a separate latent icosahedral mesh. The latent mesh and the graph connectivity are independently specified, rather than determined by the data grid. We briefly describe them here and refer to \autoref{app:ml-components} for additional details.

Throughout the training stages described in \autoref{sec:curriculum}, the following graph characteristics are invariant: graph edges carry two geometric attributes, namely normalised edge length and normalised edge direction; the processor uses 1-hop multi-scale edges, with each node directly connected to all its 1-hop neighbours at the various levels of icosahedron refinement, allowing message passing to capture both fine-scale local interactions and larger-scale spatial dependencies within a single processor step; the decoder connects each data node to its three nearest hidden nodes.

Conversely, the following graph characteristics change during the training curriculum. During global pre-training, the hidden mesh is a uniform global triangular icosahedral mesh with seven levels of refinement, and the encoder connects each hidden node to its eight nearest data nodes. The temporal downscaler's encoder, instead, uses radial cutoff edges, connecting nodes within a fixed physical distance. During stretched-grid training and fine-tuning, the hidden mesh is replaced with a triangular icosahedral mesh that has 7 and 10 levels of refinement globally and on the regional domain, respectively. In this stage, the encoder uses radial cutoff edges for both forecaster and downscaler. This allows the neighbourhood size to adapt naturally to the varying node density across the resolution transition, providing denser connectivity in high-resolution regions and sparser connectivity where the grid is coarser.


\begin{figure}[!htbp]
    \centering
    \includegraphics[scale=1]{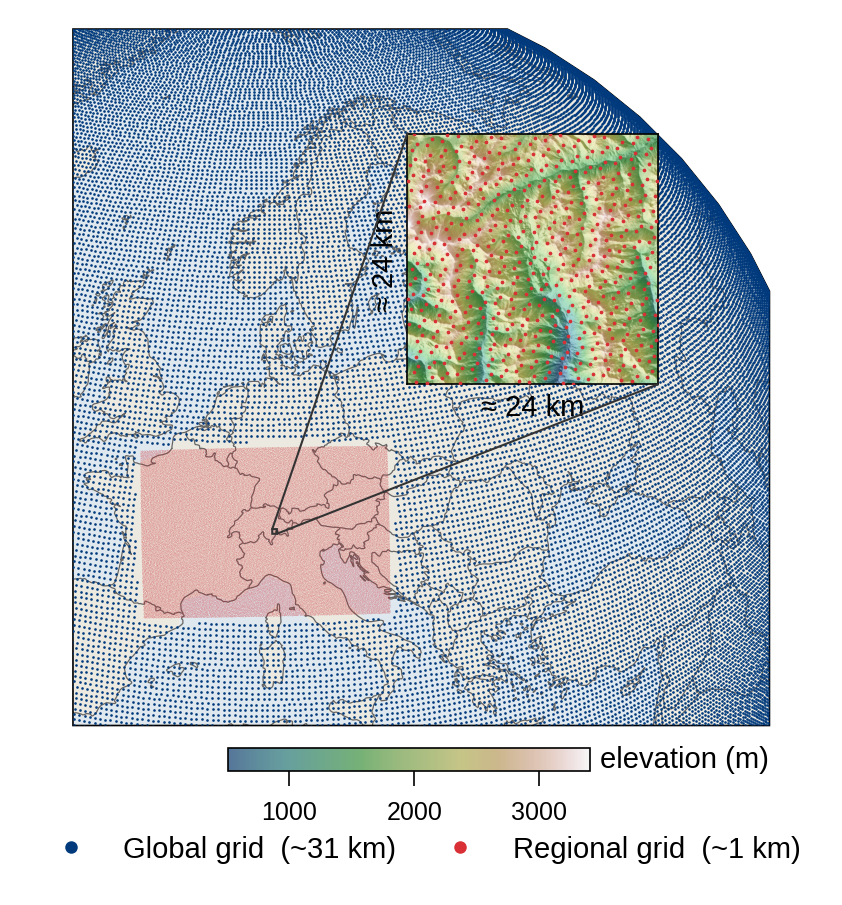}
    \caption{
    Input meshes of Varda-single. Blue nodes are obtained from the native ERA5 N320 reduced Gaussian grid and represent the spatial boundary conditions necessary to model the regional domain. Red nodes are the cell centres of the REA-L-CH1 grid at 1\,\unit{km} spatial resolution, our regional domain of interest. During pre-training, only global nodes are considered to learn global weather dynamics. In the stretched-grid training step, the regional domain is embedded within the global grid, and regional dynamics are learned conditional on the global dynamics. We display a close-up of a 24\,\unit{km} $\times$ 24\,\unit{km} cutout to depict the complex topography of the Alpine region, which spans a substantial portion of the regional domain.}
    \label{fig:meshes}
\end{figure}

\subsection{Model architecture}
\label{sec:architecture}

The forecasting and temporal downscaling models of Varda-single use the same encoder--processor--decoder architecture, with all three components using Graph Transformers as message passing layers \citep{shi2021}. 
A single encoder layer maps data node features to the latent hidden space via cross-attention over encoder edges, using 16 attention heads and a multi-layer perceptron with an expansion ratio of 4. 
A stack of 16 processor layers refines hidden-node representations of size 1024 through self-attention over the processor edges (16 heads, MLP ratio 4). The 16-layer depth combined with multi-scale processor edges gives the processor a wide receptive field spanning both the fine Alpine domain and the global circulation. 
A single decoder layer maps updated hidden representations back to data space via cross-attention over decoder edges (16 heads, MLP ratio 4). 
A more in-depth description of graph connectivity can be found in \autoref{app:graph-structure} alongside an illustrative graph visualisation in \autoref{fig:connectivity}.
For prognostic variables, a residual connection adds the most recent input state to the decoder output, making the model predict increments rather than absolute states. This stabilises training and reduces the smoothing effect. The temporal downscaler, instead, has a residual connection on the first input. A rectified linear unit (ReLU) enforces non-negativity on total precipitation \citep{moldovan2026gmd}.
The forecasting model takes the current atmospheric state and the state 6 hours prior as input, concatenating them along the feature dimension, and predicts the updated +6\,\unit{h} atmospheric state. Longer forecasts are obtained by autoregressive rollout.
The temporal downscaling model instead concatenates the bounding states of a 6\,\unit{h} time window, and jointly decodes the hourly atmospheric states in a single forward pass.

Each input state comprises the prognostic variables listed in \autoref{tab:variables} together with a set of \emph{forcing} variables that are provided as additional input but never predicted.
Forcings include trigonometric encodings of latitude, longitude, Julian day, and local solar time, top-of-atmosphere insolation, the land-sea mask, and the resolved surface orography. Two \emph{subgrid orography} fields are also included as forcings for the forecaster: the subgrid-scale standard deviation of orography and the average orographic slope. These encode terrain roughness at scales finer than the grid spacing.

Normalisation is variable-specific: prognostic variables use mean--standard deviation normalisation; total precipitation uses standard deviation only (preserving non-negativity); orographic fields are normalised by their maximum; and forcing variables with inherently bounded ranges (trigonometric encodings, insolation, land-sea mask) are passed through unchanged.

\begin{table*}[htbp]
\centering
\setlength{\tabcolsep}{3pt}
\renewcommand{\arraystretch}{1.12}
\caption{
Curriculum training stages for the 6-hourly forecaster and the hourly
temporal downscaler. LR denotes the peak learning rate. For the forecaster,
a cosine schedule with 10 warm-up steps is used. For the temporal downscaler,
the learning rate is linearly warmed up over 100 steps and subsequently
annealed with a cosine schedule to $3\times10^{-7}$ at the final step.
The effective number of samples seen by the model is the number of steps
multiplied by the global batch size.
}
\label{tab:training_stages}











\small
\setlength{\tabcolsep}{3.5pt}
\begin{tabular}{lcccccccc}
\toprule
\multicolumn{1}{l}{Training}
& \multicolumn{1}{c}{Global}
& \multicolumn{1}{c}{Regional}
& \multicolumn{1}{c}{Training}
& \multicolumn{1}{c}{Steps}
& \multicolumn{1}{c}{GPUs}
& \multicolumn{1}{c}{Global}
& \multicolumn{1}{c}{LR}
& \multicolumn{1}{c}{Rollout}\\
\multicolumn{1}{l}{stage}
& \multicolumn{1}{c}{dataset}
& \multicolumn{1}{c}{dataset}
& \multicolumn{1}{c}{period}
&
&
& \multicolumn{1}{c}{batch size}
&
& \multicolumn{1}{c}{length}\\
\midrule

\multicolumn{9}{l}{\textbf{6-hour forecaster}} \\
\midrule

Global pre-training
& ERA5 N320
& --
& 1979--2023 (45 yr)
& 50{,}000
& 32
& 32
& $10^{-5}$
& 1\\

Stretched-grid training
& ERA5 N320
& REA-L-CH1
& 2005--2023 (19 yr)
& 100{,}000
& 32
& 16
& $6\times10^{-6}$
& 1\\

Rollout training
& ERA5 N320
& REA-L-CH1
& 2005--2023 (19 yr)
& 8{,}000
& 64
& 16
& $6\times10^{-6}$
& 2--6\\

Operational fine-tuning
& IFS N320
& KENDA-CH1
& 2024--2025 (2 yr)
& 1{,}000
& 64
& 16
& $10^{-6}$
& 6\\

\midrule
\multicolumn{9}{l}{\textbf{1-hour temporal downscaler}} \\
\midrule

Global pre-training
& ERA5 N320
& --
& 1979--2023 (45 yr)
& 50{,}000
& 64
& 32
& $10^{-3}$
& 1\\

Stretched-grid training
& ERA5 N320
& REA-L-CH1
& 2005--2023 (19 yr)
& 50{,}000
& 64
& 16
& $2.5\times10^{-4}$
& 1\\

\bottomrule
\end{tabular}
\end{table*}

\subsection{Training task and objective}
\label{sec:loss}

The data-driven model minimises a variable-weighted mean squared error (MSE) loss.
Pressure-level variables receive a height-dependent multiplier that increases linearly with pressure, assigning greater loss emphasis to lower-tropospheric levels, with a minimum weight floor applied to upper-tropospheric variables.
Because the Alpine domain covers only a small fraction of the globe's surface, naive area-weighted loss assigns it a correspondingly negligible share of the total gradient signal.
A node reweighting scaler corrects this imbalance by rescaling the Alpine nodes so that their combined loss contribution equals 5\unit{\%} of the total, a significant increase above their natural area fraction.
This reweighting is active from the stretched-grid stage onward.

The temporal downscaler shares the same architecture as the 6-hourly forecaster, but they differ in the training task:
the downscaler is conditioned on two bounding atmospheric states at time $t$ and $t + 6$\,\unit{h} and jointly predicts the six hourly states from $t{+}1$\,\unit{h} through $t{+}6$\,\unit{h} in a single forward pass. 
Predicting the sixth state ($t{+}6$\,\unit{h}) is necessary to ensure the continuity of accumulated precipitation beyond the 1 hour temporal resolution. Therefore, in order to avoid inconsistencies, the $t{+}6$\,\unit{h} forecaster prediction for precipitation is overwritten with the temporal downscaler output.
The downscaler is trained independently on reanalysis data only, using the mean squared error loss with pressure-level, variable, node-weight scaling and uniform time-step scaling. The forecaster samples the training dataset with strides of 6\,\unit{h}, while the temporal downscaler samples it at hourly intervals.

\subsection{Training curriculum}
\label{sec:curriculum}
Varda-single is trained through a four-stage curriculum adapted from \citet{nipen2025aies}, described below. We outline in detail the curriculum for training the forecasting model, noting that the temporal downscaling model follows an analogous setup.
All stages use AdamW \citep{loshchilov_decoupled_2019} ($\beta_1 = 0.9$, $\beta_2 = 0.95$) with a cosine learning-rate schedule, 10 warm-up steps for the forecaster and 100 for the temporal downscaler, gradient clipping by value at 32.0 for the forecaster and by norm at 1.0 for the temporal downscaler, and 16-bit mixed precision, on up to 64 GPUs on the Alps Supercomputer (Swiss National Supercomputing Centre) using either NVIDIA A100s or NVIDIA GH200 superchips. Development and experimentation were primarily carried out on the LUMI supercomputer, using up to 128 AMD Instinct MI250X GPUs.

In the first stage of the curriculum, the model is trained on the ERA5 N320 reanalysis (31\,\unit{km}, 6-hourly, 1979--2023) using the global uniform-resolution graph.
Training data covers the years 1979--2023, while the validation period is 2024.
The objective is to acquire a broad prior over global atmospheric dynamics before any regional specialisation.

In the second stage, the global ERA5 pre-trained weights are then transferred and the graph is replaced with the stretched-grid configuration as shown in \autoref{fig:meshes}.
The training dataset is a cutout combining the REA-L-CH1 regional data with the ERA5 global data, aligned on a common 6-hourly time axis.
Training data covers the years 2005--2023, while the validation period is 2024.
This is also the stage at which the regional node reweighting described in \autoref{sec:loss} becomes active, amplifying the Alpine domain's loss contribution from its approximately 1\unit{\%} area fraction to 5\unit{\%} of the total.

In the third stage, the forecasting model is further trained with autoregressive rollout to reduce the error accumulation over longer forecasting horizons. The rollout stage begins with a rollout window of two, increasing by one at each epoch until reaching a maximum of six.

Finally, the forecasting model is fine-tuned on a cutout of KENDA-CH1 operational analyses at 1\,\unit{km} \citep{schraff2016qjrms} and IFS operational analyses at N320 for the period 2024--2025, following the approach of \citet{lang2024}.
Training data includes the year 2024, while the validation period is 01.01.2025--01.04.2025. We hold out almost a full year for testing: 01.04.2025--01.04.2026 as described in \autoref{sec:evaluation}.
Because KENDA-CH1 provides the initial conditions used at operational inference time, this stage closes the training--inference distribution gap between REA-L-CH1, which does not implement fully fledged data assimilation, and the operational analysis.
Despite its brevity, this stage produces the single largest skill improvement of the whole curriculum: RMSE and bias both distinctly decrease for many surface variables (e.g., we see a 25\unit{\%} reduction in the RMSE of 2\,\unit{m} temperature at +60\,\unit{h} lead time). Such a significant gain indicates that the model already has the necessary representational capacity and that the reanalysis-to-analysis distribution shift is the dominant residual error source. Viewed another way, the outsized effect of this stage is itself a symptom of how much the REA-L-CH1 reanalysis differs from the operational KENDA-CH1 analysis: a 20-year reanalysis is not a drop-in proxy for the operational analysis the model is initialised from, and the residual gap must be corrected explicitly. The resulting fine-tuned model is the final Varda-single configuration. 

Global operational analyses are not available at hourly resolution. For this reason, the temporal downscaler does not undergo fine-tuning and remains trained on reanalysis data only. One alternative would be to use operational forecasts for the global domain instead; we leave this exploration to future work.

\autoref{tab:training_stages} summarises the curriculum training stages for both the 6-hourly forecaster and the hourly temporal downscaler while providing detailed information about the learning rate, number of training steps and effective batch size.

\begin{figure*}[h]
    \centering
        \includegraphics[scale=1, trim=55 60 0 0]{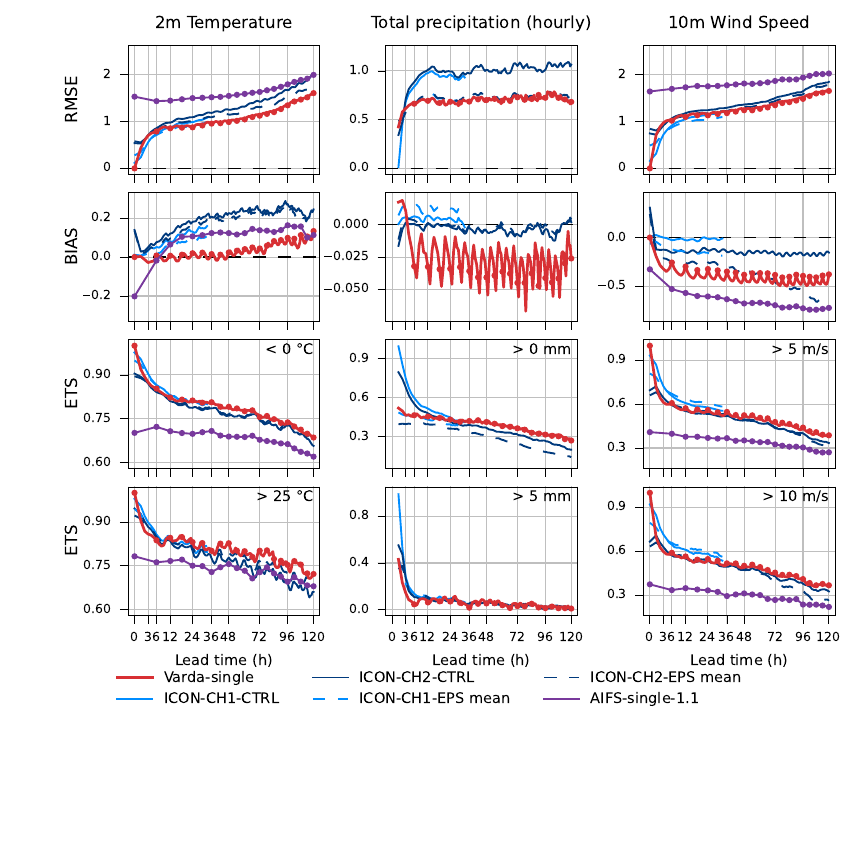}        
    \caption{Forecast performance by forecast lead time for Varda-single and the operational baseline forecasts from ICON and AIFS v1.1. RMSE (top row), bias (second row) and ETS (bottom rows) have been computed against the operational KENDA-CH1 analysis and averaged across all initialisations in the test period and all grid points in the inner verification domain. Forecasts from the 6-hourly forecaster are indicated with dots.}
    \label{fig:leadtime}
\end{figure*}

\section{Evaluation and discussion}\label{sec:evaluation}

In order to assess the performance of Varda-single, we produce forecasts initialised every 30\,\unit{h} for the verification period from April 2025 to the end of March 2026.
We evaluate hourly forecast output against both the operational analysis (KENDA-CH1 as described in \autoref{sec:datasets}) and station observations from the automatic measuring network of MeteoSwiss \citep[SwissMetNet,][also see \autoref{fig:geography-stations}]{swissmetnet2026}. As NWP baselines, the ICON-CH1-EPS and ICON-CH2-EPS models are used, and AIFS-single v1.1.0 \citep{moldovan2026gmd} is also considered as an operational data-driven baseline. ICON-CH2-EPS and AIFS-single models are interpolated to the target resolution by taking the nearest grid cell in geographical space.
ICON-CH1-EPS runs every 3 hours, producing an 11-member ensemble at 1\,\unit{km} horizontal resolution up to +33\,\unit{h} lead time. ICON-CH2-EPS runs every 6 hours, producing a 21-member ensemble at 2\,\unit{km} horizontal resolution up to +120\,\unit{h}.
We use both the unperturbed control simulation and the ensemble mean from the operational NWP forecasts as baselines. They are referred to as ICON-CH1-CTRL and ICON-CH1-EPS mean, respectively. 
AIFS-single v1.1.0 runs operationally every 6\,\unit{h} at approximately 31\,\unit{km} resolution out to 15 days ahead; here we use forecasts up to +120\,\unit{h}. 
Including AIFS-single, developed as a global forecasting system rather than a kilometre-scale regional model, provides a data-driven baseline to show the added value of regional modelling.
Before further evaluation, we regrid the forecasts to the grid of the reference dataset, and for temperature we apply a constant lapse-rate correction of 0.0065\,\unit{K\,m^{-1}} to account for altitude differences. 
First, we show aggregated verification scores and briefly discuss the overall patterns in the error distribution. 
We then focus on case studies that are representative of the current strengths and limitations of Varda-single.

\subsection{Overall model performance}
\label{sec:overall_scores}

\begin{figure*}[!htp]
    \centering
        \includegraphics[scale=1, trim = 0 0 0 25, clip]{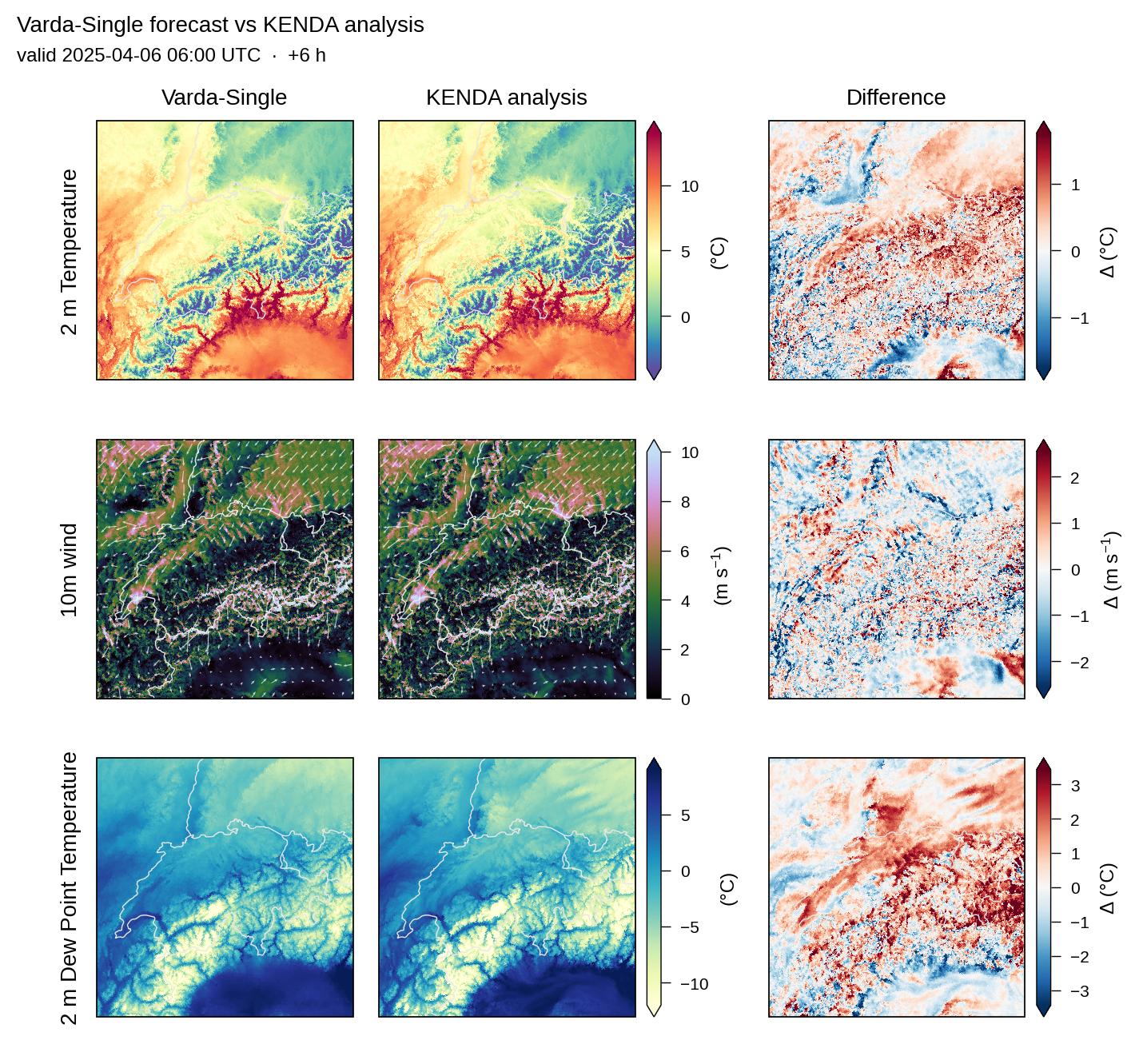}
    \caption{An example of a single 6-hourly forecasting step of Varda-single initialised at 06~UTC on 6 April 2026. The left column shows the model output for 2\,\unit{m} temperature, 10\,\unit{m} wind and 2\,\unit{m} dew point temperature; the middle column shows the corresponding operational KENDA-CH1 analysis. The right column shows the pixel-wise difference between the forecast and the analysis.}
    \label{fig:teaser-comparison}
\end{figure*}

We compare Varda-single, alongside ICON-CH1-EPS, ICON-CH2-EPS and AIFS-single v1.1.0, to MeteoSwiss' operational analysis (KENDA-CH1).
For model evaluation, we compute the mean bias (BIAS), the root-mean-squared error (RMSE), and the equitable threat score (ETS) for binary threshold-exceedance events. 
Those metrics are shown in \autoref{fig:leadtime} for 2\,\unit{m} temperature, hourly total precipitation, and 10\,\unit{m} wind speed.

For temperature, Varda-single is slightly worse in terms of RMSE than the high-resolution operational baseline up to +12\,\unit{h}; thereafter, performance is comparable to the performance of the NWP ensemble mean.
Varda-single exhibits a consistently smaller bias than all operational baselines shown here, especially early in the forecast when the bias is negligible. 
The near-zero bias at short lead times follows from the KENDA-CH1 fine-tuning, which aligns Varda-single's climatology with the operational analysis.
Varda-single's threshold exceedance forecasts for temperature are on par with or slightly better than the operational NWP baselines apart from the first few hours of the forecast, as illustrated by the ETS for 2\,\unit{m} temperature below 0\,\unit{\degree C} and above 25\,\unit{\degree C}.

For total precipitation, Varda-single matches the RMSE of the operational NWP ensemble means for the available forecast ranges, and outperforms the unperturbed members. 
After +3\,\unit{h}, Varda-single shows a larger negative bias than the NWP models. 
This pattern is consistent with the generally smoother precipitation produced by data-driven models \citep{subich2025}, which achieve competitive RMSE values but underestimate high-intensity precipitation. 
The ETS for distinguishing between dry and wet hours (total precipitation > 0\,\unit{mm}) is higher for Varda-single than for the NWP baselines after +24\,\unit{h} lead time, whereas before, it is outperformed by the unperturbed NWP members. 
For thresholds above 5\,\unit{mm}, ETS is comparable to or slightly worse than that of the NWP baselines early in the forecast. 

For 10\,\unit{m} wind speed, Varda-single and NWP show a similar RMSE, with Varda-single performing slightly better than ICON-CH2-EPS (control and mean), and slightly worse than ICON-CH1-EPS (control and mean). 
Varda-single's ETS matches that of ICON-CH2-EPS (control and mean) for both analysed thresholds for lead times up to +48\,\unit{h}, and slightly exceeds it at longer lead times. 
The high-resolution NWP baseline outperforms both Varda-single and the lower resolution ICON model in terms of ETS.

Hourly precipitation is not available from the 6-hourly AIFS v1.1 forecaster and thus cannot be compared.
For near-surface temperature and wind speed, we find that Varda-single consistently outperforms AIFS v1.1 in terms of RMSE and ETS throughout the forecast range.
The rate of error growth, however, is somewhat lower in AIFS, resulting in a smaller benefit of high-resolution forecasting at longer lead times.
As the two data-driven systems are very similar, the comparison of Varda-single and AIFS v1.1 mostly highlights the expected benefit of high-resolution data-driven forecasting in complex topography and we refrain from showing AIFS v1.1 in the remainder of the paper. 

\begin{figure*}[!t]
    \centering
    \includegraphics[scale=1]{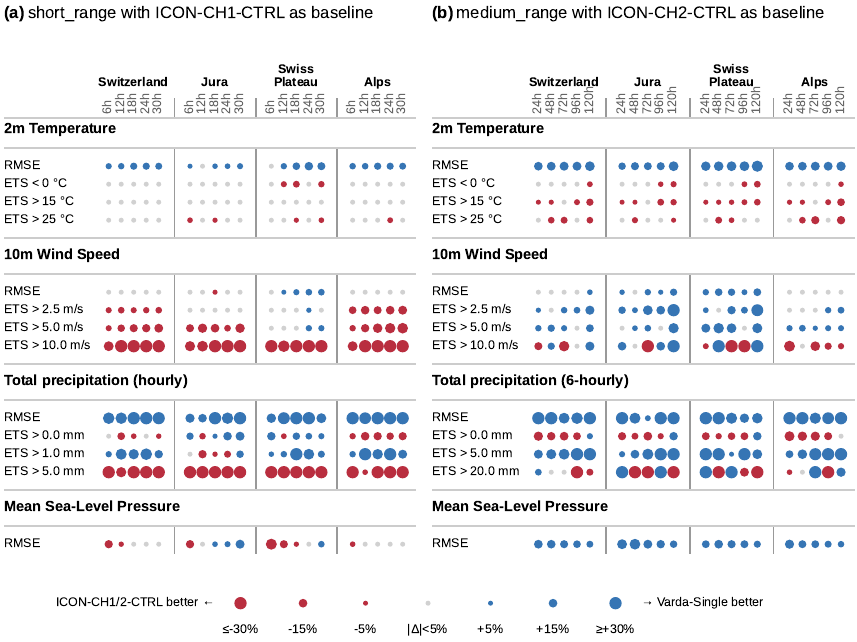}
    \caption{Station-based scorecards: relative percentage difference for RMSE and ETS comparing Varda-single with ICON-CH1-CTRL (panel a) and ICON-CH2-CTRL (panel b). Blue dots: Varda-single better than the baseline NWP. Red dots: NWP baseline better than Varda-single. The size of the dots scales with the magnitude of the relative difference. }
    \label{fig:scorecard}
\end{figure*}

In addition to the analysis-based evaluation, we also evaluate forecasts against SwissMetNet station observations. 
We show scorecards of forecast quality using ICON-CH1-CTRL and ICON-CH2-CTRL as baselines in \autoref{fig:scorecard}. 
We compute the relative percentage difference against the respective baselines for RMSE and ETS of 2\,\unit{m} temperature, 10\,\unit{m} wind, hourly/6-hourly total precipitation and mean sea-level pressure.
The results are stratified by lead time, ETS thresholds, and four regions: the whole of Switzerland, the hilly Jura region in western Switzerland, the mountainous region of the Swiss Alps, and the mostly flat region between the Jura Mountains and the Swiss Alps known as the Swiss Plateau (see \autoref{fig:geography-stations} for a short description of those regions).

The forecast quality of Varda-single as measured by the RMSE is comparable to or higher than that of the high-resolution baseline ICON-CH1-CTRL (\autoref{fig:scorecard}a). 
Some of the skill as measured by RMSE is a direct consequence of training on MSE and the resulting smoothing of forecast fields. 
This effect is expected to be most notable in cases of limited predictability and explains the pronounced improvement in RMSE of hourly precipitation.
In contrast, Varda-single performs worse than ICON-CH1-CTRL in predicting relatively rare events as illustrated by the ETS for high thresholds of wind speed and precipitation.
Also, there is some indication that Varda-single performs worse in complex topography as illustrated by the ETS for wind speed. Its ETS over the Swiss Plateau is comparable to that of ICON-CH1-CTRL for thresholds up to 2.5\,\unit{m\,s^{-1}}, whereas it underperforms in the Jura Mountains and Swiss Alps.
For mean sea-level pressure, we see some indication of underperformance early in the forecast, whereas the performance as measured by RMSE is comparable after +12\,\unit{h}.

Compared against the lower resolution ICON-CH2-CTRL (\autoref{fig:scorecard}b), Varda-single shows mostly comparable or better RMSE and ETS. 
The largest improvement over the baseline is visible in the 2\,\unit{m} temperature, hourly total precipitation and mean sea-level pressure RMSE. 
Varda-single's skill is retained over the whole lead time range available from the NWP forecast.

In summary, Varda-single performs as well as or better than the operational baselines with some notable exceptions, described at length in \autoref{sec:conclusions}, which will be addressed in future work. 
Mainly, deterministic data-driven models trained to minimise MSE reproduce predictable components well, but attenuate peak intensity and high-frequency variability. 
This is the expected consequence of regressing toward the conditional mean, the effect of which is evident where the atmosphere is least predictable at the resolved scale, i.e. precipitation and wind in complex terrain.
Probabilistic or ensemble formulations and spatially aware loss functions \citep{nordhagen2025, subich2025} are known to alleviate these limitations.
In addition, we observe underperformance of Varda-single in the first few hours of the forecast.
To better understand the capabilities and limitations of the current Varda model's forecasts, we present an analysis of spatial variability of forecast performance in \autoref{sec:spatial-skill} and three case studies in \autoref{sec:case-studies}. 

\subsection{Spatial distribution of forecast skill}
\label{sec:spatial-skill}

In the following we present an analysis of the spatial distribution of forecast skill, pointwise rather than as a regional average.
Metrics are accumulated over all initialisation times and evaluated at every grid point $\mathrm{x}$ of the verifying KENDA-CH1 analysis, at a fixed lead time.
We use the MSE skill score $S(\mathrm{x}) \in (-\infty, 1]$ of the forecaster ($\mathrm{f}$) against the ICON-CH1-CTRL baseline ($\mathrm{b}$), which, by splitting the mean-square error into squared bias and error variance, decomposes additively into a systematic ($S_\mathrm{BIAS}$) and a random ($S_\mathrm{STDE}$) contribution,

\begin{equation}
\begin{split}
S(\mathrm{x}) &= 1 - \frac{\mathrm{MSE_f}}{\mathrm{MSE_b}} \\[2pt]
&= \underbrace{\frac{\mathrm{BIAS_b^2}-\mathrm{BIAS_f^2}}
                    {\mathrm{MSE_b}}}_{S_\mathrm{BIAS}}
 + \underbrace{\frac{\mathrm{STDE_b^2}-\mathrm{STDE_f^2}}
                    {\mathrm{MSE_b}}}_{S_\mathrm{STDE}} .
\end{split}
\label{eq:skill-decomposition}
\end{equation}

We note that the terms ``random'' and ``systematic'' may be slightly misleading here, since the random component may include systematic components that are not revealed in this analysis, and would only appear under further stratification, for example by time of day, valid time, initialisation time, or more detailed topographic characteristics. 

\begin{figure}[!htbp]
    \centering
    \includegraphics[scale=1]{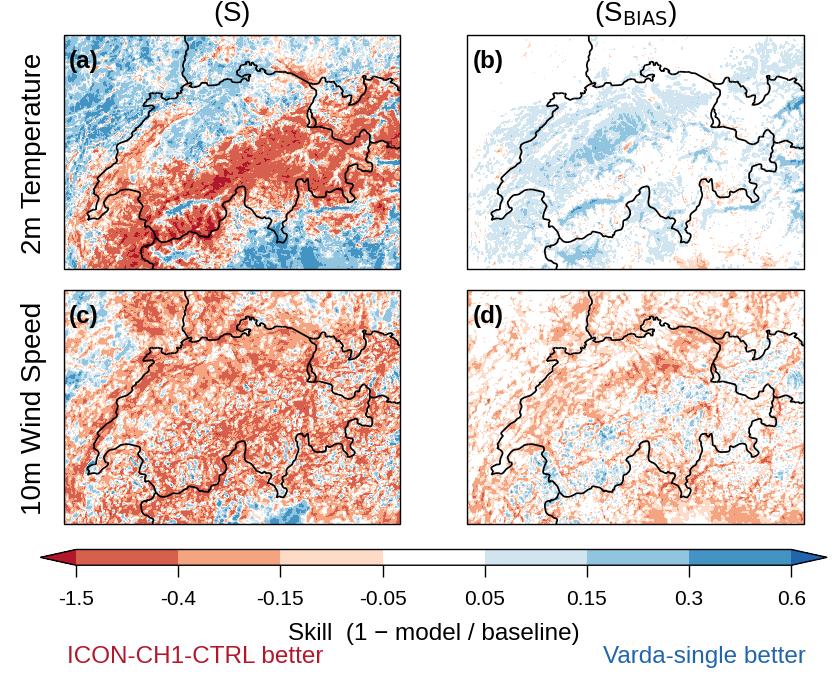}
    \caption{Pointwise MSE skill score of Varda-single against the ICON-CH1-CTRL
    baseline at +6\,h lead time, accumulated over all initialisations and
    verified against the KENDA-CH1 analysis. Top row: 2\,\unit{m} temperature; bottom row:
    10\,\unit{m} wind speed. Left column: total skill score $S$
    (\autoref{eq:skill-decomposition}). Right column: the contribution
    $S_\mathrm{BIAS}$ of the systematic error; the random contribution
    $S_\mathrm{STDE}$ is the difference between the two columns. Blue indicates that
    Varda-single outperforms the baseline, red that it underperforms. Both columns
    share the colour scale, which saturates beyond $\pm0.65$.}
    \label{fig:scoremaps_6h}
\end{figure}

\autoref{fig:scoremaps_6h} shows $S$ (left column) and $S_\mathrm{BIAS}$ (right column) for 2\,\unit{m} temperature and 10\,\unit{m} wind speed over Switzerland and the surrounding Alpine region. 
Both $S$ and $S_\mathrm{BIAS}$ are dimensionless and $S_\mathrm{STDE}$ can be read off as the difference between the two.
Two properties of this decomposition matter for the interpretation.
First, $S_\mathrm{BIAS}$ compares the forecaster's systematic error with that of the baseline: a value near zero means that both systems are similarly biased at that point, not that either is unbiased.
Second, $S$ is unbounded below, so the magnitude of the negative skill is locally larger than the colour scale suggests.
We show the maps at +6\,h lead time, i.e.\ after a single autoregressive step of the forecaster, to isolate the spatial signal without the confounding influence of rollout error. 

For 2\,\unit{m} temperature (\autoref{fig:scoremaps_6h}a), Varda-single's performance is quite heterogeneous but shows a clear spatial signal: positive skill is observed over low-lying flat regions and major mountain valleys, and negative skill is mostly found over major lakes and high-elevation regions, especially where the terrain gradients are very steep. \autoref{fig:scoremaps_6h}b allows us to distinguish the systematic and random contributions towards MSE skill for 2\,\unit{m} temperature. 
Overall, $S_\mathrm{BIAS}$ is small compared to the MSE skill score in \autoref{fig:scoremaps_6h}a. 
Thus, the contribution towards overall skill seems to be predominantly random in nature. 
This is especially true at high elevations, where the overall forecaster skill is negative. 
However, the systematic component is also small in some areas of low elevation (in the north and south-east of the domain shown), and hence the random contribution to overall positive skill is large there. 
Some regions, such as the floors of major mountain valleys and large parts of the Swiss Plateau, show substantial systematic contributions to overall skill. 
In the case of the mountain valleys, we hypothesise that the Varda-single forecaster shows a smaller warm bias than the ICON-CH1-CTRL baseline, which systematically underestimates cold-air pool amplitude.
We do not have a clear explanation for the systematic contribution to positive skill in the Swiss Plateau. 

For wind speed (\autoref{fig:scoremaps_6h}c), Varda-single's overall skill is negative throughout most of the domain shown. 
Exceptions are small patches with positive values in the mountainous areas, and some larger areas in the flat lands in the north-west and the south of the depicted domain. 
The split of total skill into squared bias (\autoref{fig:scoremaps_6h}d) and error variance (not shown) reveals the following:  The bias contribution to overall skill is negative or neutral throughout most of the domain. 
The only exceptions are some high-mountain locations, where the bias contribution is slightly positive. 
The bias maps (not shown) indicate that Varda-single's bias is negative throughout most of the domain, whereas the bias of ICON-CH1-CTRL is neutral at low elevations and negative at high elevations. 
At low elevations, Varda-single therefore underperforms ICON-CH1-CTRL due to its negative bias, whereas it outperforms it in some high-mountain locations because its negative bias there is smaller. 
Varda-single's negative wind-speed bias may in part be explained by the fact that the model is trained to predict the two wind components. 
The MSE training objective introduces a conditional bias in both components, rewarding smaller absolute values in general. 
This bias in wind components is then transferred and amplified when wind speed is computed. 

The error variance contribution to overall skill (difference of panels c and d of \autoref{fig:scoremaps_6h}, not shown) is negative throughout most of the domain, without a clear spatial pattern. 
The standard deviation maps of the two models, however, are instructive: 
Both models show a similar pattern of lower standard deviation at low elevations, and higher standard deviation at higher elevations. Varda-single's standard deviation is higher overall, leading to a negative skill contribution. 
This contribution is particularly pronounced at high elevations, whereas the signal at low elevations is weaker and more mixed. 
This gives some indication that it is challenging for Varda-single to represent wind phenomena (such as localised channelling, blocking and gap-flow dynamics) in complex topography, leading to a large random error. 

We note that the scorecard in \autoref{fig:scorecard}a and the maps in \autoref{fig:scoremaps_6h}a,c show some apparent disagreement:
For 2\,\unit{m} temperature, the scorecard shows Varda-single outperforming ICON-CH1-CTRL in terms of RMSE, whereas \autoref{fig:scoremaps_6h}a shows the opposite, particularly over the Alps. 
For 10\,\unit{m} wind speed, the scorecard shows similar performance of both models over Switzerland and the Alps, whereas the map in \autoref{fig:scoremaps_6h}c shows mostly better ICON-CH1-CTRL skill. 
We attribute this apparent discrepancy to the different reference datasets used:
The scorecard uses station observations, whereas the maps use the KENDA-CH1 analysis.

\subsection{Case studies}
\label{sec:case-studies}
The following section provides a critical assessment of Varda-single through three emblematic case studies, chosen to investigate its behaviour during significant weather and thus complement the large-sample analyses presented above. We first examine thermally driven valley winds to assess the model's ability to represent terrain-driven local circulations. We then consider a Foehn event to distinguish the representation of large-scale forcing from the small-scale processes that govern flow near complex terrain. Finally, we examine precipitation in both synoptic and convective regimes, focusing on the spatial structure of the forecast field and the limitations imposed by the deterministic training objective. To give a visual impression of Varda-single forecasts, we provide \autoref{fig:teaser-comparison} showing a 6-hourly Varda-single forecast for 2\,\unit{m} temperature, 10\,\unit{m} wind and 2\,\unit{m} dew point temperature, compared to the corresponding operational KENDA-CH1 analysis.

\subsubsection{Thermally driven valley winds in steep terrain}
\label{sec:case-valley}

\begin{figure*}[h]
    \centering
    \includegraphics[scale=1]{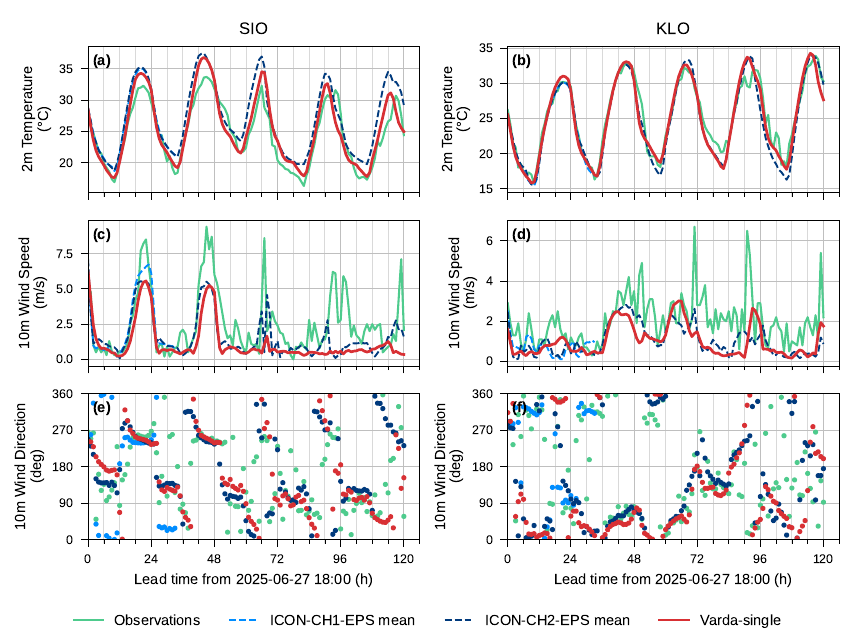}
    \caption{Meteograms for the forecast initialised at 18\,UTC on 27~June~2025 for 2\,\unit{m} temperature and 10\,\unit{m} wind speed and direction. The left column shows the valley station in Sion (SIO) and the right column the flat-region reference in Kloten (KLO), for SwissMetNet station observations (green), ICON-CH1-EPS (light blue) and ICON-CH2-EPS (dark blue) ensemble means, and Varda-single (red).}
    \label{fig:case-valley}
\end{figure*}

The aggregated scorecards and skill maps (\autoref{sec:spatial-skill}) indicate that mountain weather is difficult for Varda-single to predict. To further investigate this, we present a process-level case study with a focus on wind at two contrasting stations: Sion (SIO), located on the floor of a major Alpine valley oriented east-west with a number of side valleys (the Valais), and Z\"urich-Kloten (KLO) on the flat Swiss Plateau, see \autoref{fig:geography-stations}. \autoref{fig:case-valley} shows observed 2\,\unit{m} temperature and 10\,\unit{m} wind speed and direction at these stations, together with forecasts initialised at 18:00\,UTC on 27~June~2025 from the operational ICON-CH1-EPS and ICON-CH2-EPS ensemble means and Varda-single. 

The period shown splits into two contrasting regimes at the SIO station. 
The first two days (up to +54\,\unit{h} lead time) are influenced by weak synoptic forcing and a high-pressure system. 
A characteristic thermally driven valley circulation develops, with daytime up-valley winds, nocturnal down-valley (katabatic) winds, and a pronounced nighttime surface temperature inversion in the valley. 
Also, there is no precipitation in the vicinity of the valley (not shown). 
From 30~June onward (after +54\,\unit{h} lead time) the weather regime becomes convective, with several precipitation events over the region peaking near 5\,\unit{mm\,h^{-1}} around 1~July. 
These convective episodes reorganise the local flow on sub-diurnal timescales and provide a stringent test of each model's response to rapid, non-periodic forcing.

For 2\,m temperature, Varda-single reproduces the diurnal cycle at both stations
with high fidelity throughout the forecast and outperforms both ICON baselines. At SIO
(\autoref{fig:case-valley}a) the ICON-CH1-EPS and the ICON-CH2-EPS
ensemble means underestimate nighttime cooling and overshoot the daytime maxima -- the lower-resolution NWP mean exceeds the
observed peaks by up to approximately $3\,\mathrm{K}$ on several days -- whereas
Varda-single follows the SwissMetNet curve more closely, including the
nocturnal minima associated with valley-floor cooling under the inversion. The
same ranking holds at the flat-land station KLO (\autoref{fig:case-valley}b), though the inter-model spread is smaller there. 

The 10\,m wind speed exposes the central limitation. During the weak forcing regime
Varda-single correctly reproduces the \emph{phase} of the valley-wind cycle at SIO
(\autoref{fig:case-valley}c): both the daytime acceleration and the weak nocturnal winds
tied to the inversion are present. However, it systematically
underestimates the \emph{intensity}, rendering observed afternoon peaks of around
8\,\unit{m\,s^{-1}} as only approximately 5\,\unit{m\,s^{-1}}, while the
ICON-CH1-EPS mean retains more of the amplitude. 

During the second part of the episode, the heavy convective rainfall over the valley disturbs the diurnal build-up of the valley-wind circulation. 
This is evident in \autoref{fig:case-valley}c and \autoref{fig:case-valley}d, where the observations show much more erratic and high-frequency variability than during the calm period. 
While ICON-CH2-EPS reproduces this behaviour in wind speed to some degree (\autoref{fig:case-valley}c), Varda-single evidently does not. 
The forecaster's 6-hourly steps miss most of this high-frequency variability, and the temporal downscaler model then struggles to fill in the gaps. 
Trained using an MSE loss function, Varda-single responds to these high-frequency conditions (and associated high uncertainty) by falling back to a conservative estimate, producing only very low wind speeds in order to avoid large errors. 
This characteristic deficiency of Varda-single is also evident in the other case studies. 


\subsubsection{Foehn wind}
\label{sec:case-foehn}

In this case study we examine the behaviour of Varda-single during a typical south Foehn event, a combination of low pressure on the northern side of the Alps and high pressure in the south that drives air up the southern slope of the Alps. During ascent, the air cools, leading to cloud formation and precipitation. The now-drier air descends the lee slope, is compressed and warms rapidly, arriving in the northern valleys as a warm, dry and gusty downslope wind. 

\begin{figure*}
    \centering
    \includegraphics[scale=1]{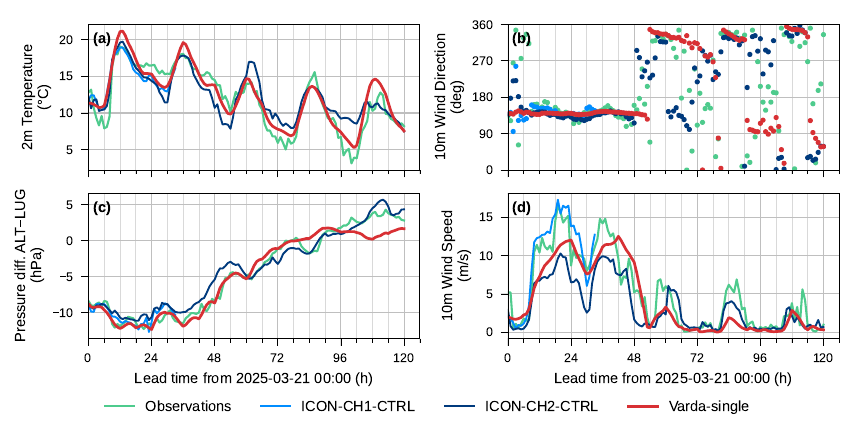}
    \caption{Meteograms for the forecast initialised
    at 00\,UTC on 21~March~2025 for 2\,\unit{m} temperature (a), 10\,\unit{m} wind direction (b) and speed (d) in Altdorf (ALT), and the mean sea-level pressure difference between ALT and Lugano (LUG, c), for SwissMetNet station observations (green), ICON-CH1-CTRL (light blue) and ICON-CH2-CTRL (dark blue), and Varda-single (red).}
    \label{fig:case-foehn}
\end{figure*}

We consider the south Foehn situation of 21 to 23 March 2025. Storm Martinho, approaching from the west, drives a southerly flow over the Alpine arc, associated with a cross-Alpine surface pressure gradient. This gradient is visible in the difference in mean sea-level pressure between the station in Lugano (LUG), on the southern side of the Alps, and the station in Altdorf (ALT), on the northern side (see \autoref{fig:geography-stations}). It is successfully reproduced by Varda-single, which is able to capture the forcing that drives the event, as apparent in \autoref{fig:case-foehn}c. 

The wind speed on the leeward side (ALT) exhibits two strong peaks during the event (evenings of 21 and 22 March, around +18\,\unit{h} and +42\,\unit{h} lead time) with a maximum of 15\,\unit{m\,s^{-1}} (\autoref{fig:case-foehn}d). Varda-single places the onset correctly and reproduces the shape and timing of both wind speed peaks, but attenuates their amplitude. The three models therefore accurately forecast timing, but differ in their ability to forecast intensity, a similar pattern to that observed for the valley winds in \autoref{sec:case-valley}. Consistent with the general behaviour observed for Varda-single, the attenuation manifests in wind speed, whose magnitude is set by sub-kilometre-scale dynamics, and not in the synoptic gradient. The ICON-CH1-CTRL forecast is able to represent the wind speed amplitude much better. 
The temperature evolution during the event (\autoref{fig:case-foehn}a) is reproduced considerably better. The arrival of the descending air (around lead time +8\,\unit{h}) causes a significant increase in 2\,\unit{m} temperature throughout the Foehn event. Varda-single accurately captures the timing of the arrival of warm air, although it considerably overestimates the midday maxima on 22 and 23 March.

The Foehn case sharpens the diagnosis of \autoref{sec:case-valley}. Varda-single is accurate for aspects of the phenomenon set by the resolved large-scale state: the forcing gradient, the onset and decay timing, the spatial asymmetry, and the several-kelvin warming are all reproduced. Conversely, for the aspects that are defined by small-scale processes and flow acceleration through terrain, the system shows lower performance. We attribute this to two compounding factors. The first is insufficient latent space computational capability due to unbalanced hidden-mesh density. Preliminary experiments outside of the scope of this paper showed an improvement in predicted wind intensities over the Alpine terrain, which we take as tentative support for the hypothesis rather than firm proof. The second is a more general property of MSE-trained models which are known to underestimate the intensity of extremes and small-scale phenomena as a result of the spectral smoothing \citep{bonavita2023}. 

Unlike the other case studies, this event falls within the validation period. We still include it for the physical insight it offers into the model's capabilities.

\subsubsection{Precipitation in synoptic and convective situations}
\label{sec:case-precip}

Precipitation is the variable for which the limitations of the pointwise deterministic training objective are most apparent. In training, the MSE loss in \autoref{sec:loss} is minimised by predicting the conditional mean of the outcome distribution. For precipitation, this conditional mean is smoother than any physically plausible realisation \citep[drizzle bias,][]{moldovan2026gmd}. In verification, pointwise scores reward this smoothness, since a sharp forecast with a slightly displaced feature scores worse than a smooth one hedging across possible locations (the double-penalty effect, \cite{ebert2008}). Varda-single is thus trained towards smoothness, and the scores in \autoref{sec:overall_scores} do not penalise it. Appropriately assessing precipitation forecast quality therefore requires metrics that measure the spatial organisation of the field rather than its values at individual grid points.

We use the SAL framework of \citet{wernli2008}, an object-based verification method that is designed to address the above concerns. It decomposes the error of an accumulated precipitation field into three components. The amplitude component $A$ measures the relative error in the domain-mean amount, the location component $L$ measures the displacement of the precipitation objects and the distribution of their mass about the domain centroid, and the structure component $S$ compares the volume-to-peak ratio of the identified objects: $S > 0$ indicates objects that are too large and too flat relative to the reference, $S < 0$ objects that are too small or too peaked. These metrics allow us to diagnose some of the characteristics of precipitation fields that simpler scores are uninformative about. We compute SAL with the \texttt{pysteps} implementation \citep{pulkkinen2019} on 6-hourly accumulations, verified against the KENDA-CH1 analysis, for lead times of +6\,\unit{h} to +30\,\unit{h}.

\begin{figure*}[!htbp]
    \centering
    \includegraphics[width=0.85\linewidth]{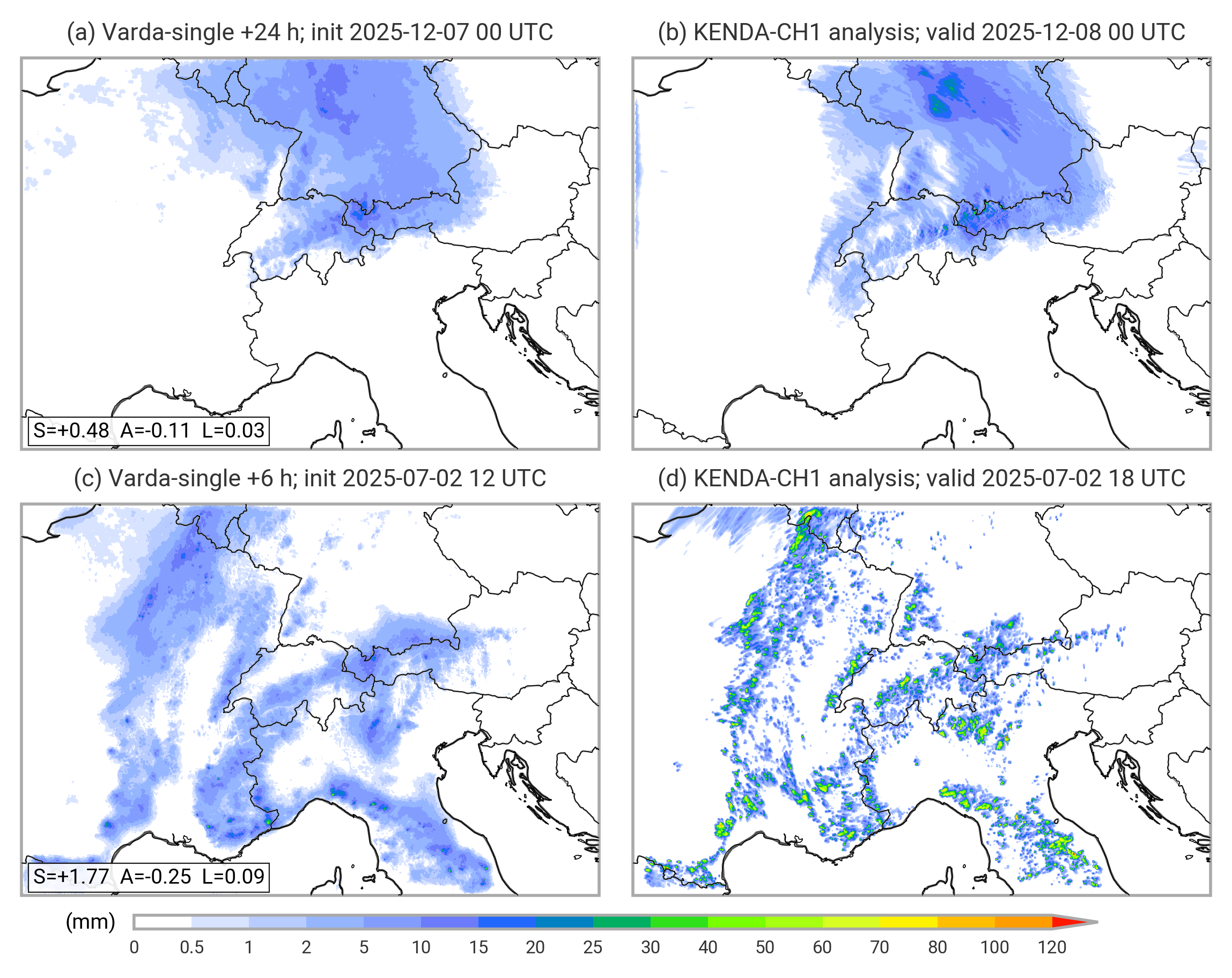}
    \caption{6\,\unit{h} accumulated precipitation for a well-forecast winter case (top; initialised at 00\,UTC on 7~December~2025, lead time +24\,\unit{h}) and a poorly forecast summer case (bottom; initialised at 12\,UTC on 2~July~2025, lead time +6\,\unit{h}). Left: Varda-single. Right: the verifying KENDA-CH1 analysis. Inset boxes give the SAL components for the displayed 6\,\unit{h} window. Note that these are single-window values, whereas the values in \autoref{fig:case-precip-scatter} are per-initialisation averages over the 6--30\,\unit{h} windows and are therefore numerically different.}
    \label{fig:case-precip-maps}
\end{figure*}

\autoref{fig:case-precip-maps} contrasts two characteristic snapshots, one from each end of the seasonal spectrum. On 7~December~2025 (\autoref{fig:case-precip-maps}a,~b), a synoptic-scale system propagates across the northern half of the domain. Varda-single and the analysis show recognisably the same weather, with only mild over-smoothing. On 2~July~2025 (\autoref{fig:case-precip-maps}c,~d), during widespread convection, the two fields show qualitatively different patterns. The analysis consists of small, intense cells reaching 143\,\unit{mm} in the 6\,\unit{h} window; Varda-single produces broad, smooth, low-intensity bands with essentially no cell structure and a domain maximum of 45\,\unit{mm}, giving $S = +1.77$, close to the maximum possible value of the score. It is worth noting that this is the \emph{first} forecasting step: the smoothing is fully developed after a single autoregressive iteration and is therefore a property of the trained one-step forecaster, not only an accumulation of rollout error. Consistent with this interpretation, the structure component barely changes across lead times, while the location component rises noticeably: structural error is established at once and then saturates, whereas displacement error grows as predictability is progressively lost (not shown). This suggests that the small-scale organisation of convective precipitation is already largely unpredictable at a range of 6\,\unit{h}. Hence, the conditional mean targeted by the MSE objective is smooth from the outset. 

\begin{figure*}[!htbp]
    \centering
    \includegraphics[width=0.88\linewidth]{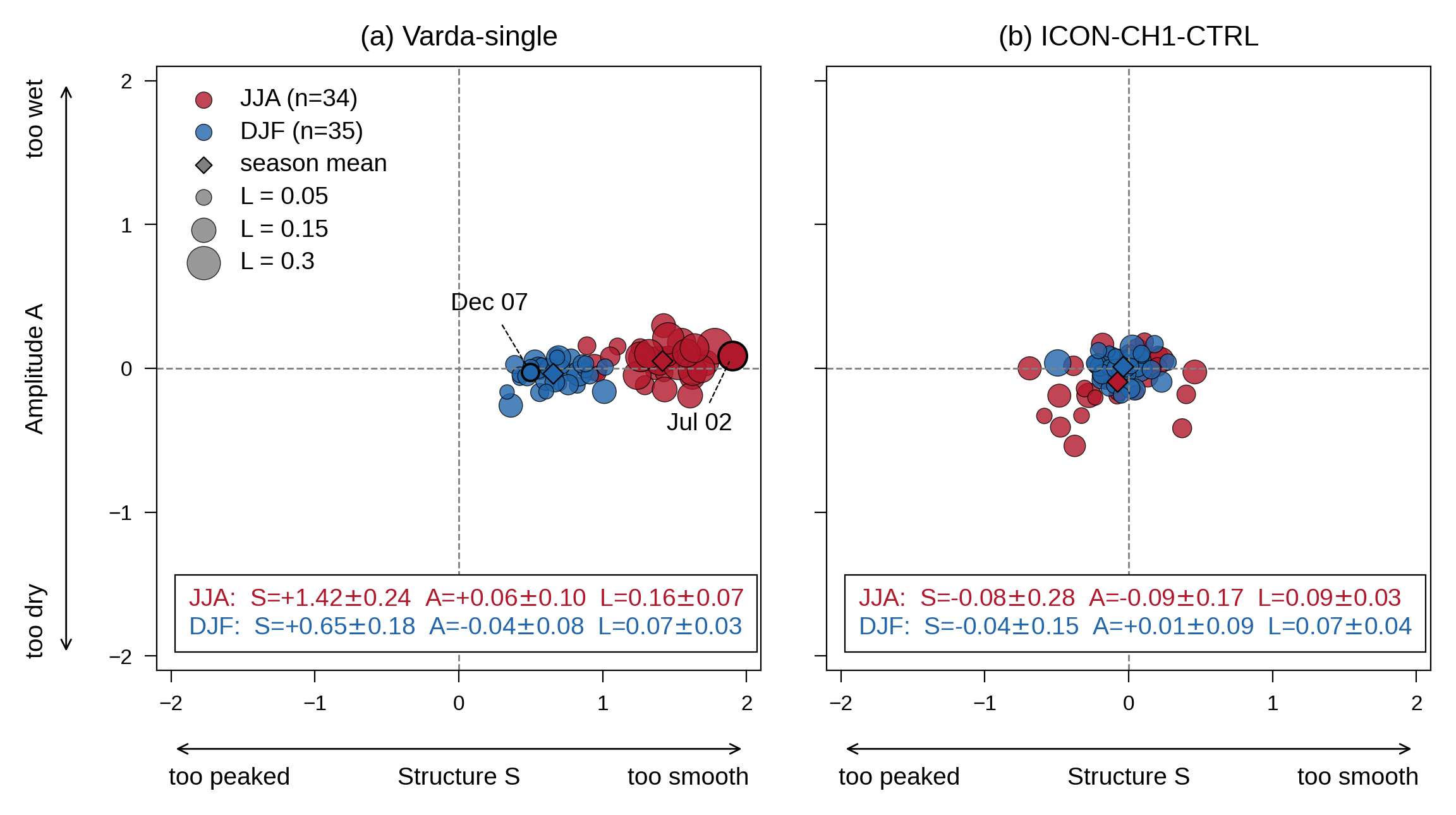}
    \caption{SAL statistics over a subset of the test period for (a) Varda-single and (b) ICON-CH1-CTRL. Each circle is one initialisation, positioned by its mean structure ($S$) and amplitude ($A$) components over the 6--30\,\unit{h} windows, with the circle area scaling with the mean location component ($L$); diamonds mark the seasonal means. Only initialisations with at least 2\,\unit{mm} of domain-mean precipitation in the KENDA-CH1 analysis over the scored window are shown (34 in JJA~2025, 35 in DJF~2025/26). The two events shown in \autoref{fig:case-precip-maps} are labelled.}
    \label{fig:case-precip-scatter}
\end{figure*}

\autoref{fig:case-precip-scatter} shows that these two cases are characteristic examples of a robust pattern. Each point is one initialisation, positioned by its mean $S$ and $A$ over the 6--30\,\unit{h} windows. Colours distinguish summer and winter cases. The two cases presented in \autoref{fig:case-precip-maps} were selected as near-extremes of a ranking derived from these statistics. For Varda-single (\autoref{fig:case-precip-scatter}a) the summer and winter strata are almost disjoint in the structure component, while the amplitude component is near zero in both seasons. Varda-single thus approximately reproduces seasonal precipitation amounts but fails to capture the spatial structure of precipitation. This is more pronounced in summer (convection) and still considerable in winter. For ICON-CH1-CTRL (\autoref{fig:case-precip-scatter}b), $S$ is close to zero in both seasons, indicating that the deficiency in structure is specific to the data-driven model. Displacement plays only a minor role, with small values of $L$ throughout. Results for ICON-CH2-CTRL are qualitatively similar to the ones from ICON-CH1-CTRL and are not shown. 

The remedy is therefore not higher resolution but a training objective that does not reward the conditional mean, as demonstrated by the CRPS-trained, spectrally constrained version of Bris \citep{nordhagen2025}. Comparable gains are reported for probabilistic regional formulations more generally \citep{larsson2025}, and for the temporal dimension of the same problem \citep{ingstad2026}.

\section{Conclusions} 
\label{sec:conclusions}
We presented Varda-single, a data-driven weather prediction system specifically built for operational use in the Alpine domain. The system is composed of two independently trained models, both fully developed within the Anemoi framework and employing the same stretched-grid Graph Transformer approach with encoder--processor--decoder architecture. The two models are the autoregressive forecaster advancing in 6-hourly steps to minimise error growth while still capturing the diurnal cycle, and a temporal downscaler that produces hourly forecasts in between the 6-hourly outputs of the forecaster. The introduction of Varda-single represents an important step in complementing the operational NWP regional models of MeteoSwiss with data-driven alternatives. It also provides a first assessment of the capabilities of regional medium-range data-driven weather prediction at 1\,\unit{km} resolution.

Verified against the operational analysis and station observations, Varda-single is competitive with or improves on the operational NWP baselines running at MeteoSwiss for most headline scores and variables.  
Our model performance analysis highlights that Varda-single performs well for average conditions but fails to capture individual realisations of the weather processes or produce sharp forecasts at the target resolution. This result is in line with our expectations: data-driven weather models trained deterministically are known to produce overly smooth forecasts. More evidence for this, beyond verification scores, can also be found in our three case studies, which demonstrate what this behaviour means in terms of strengths and weaknesses of the system.

We currently identify the following limitations of Varda-single that motivate further investigation in future work. 

The temporal downscaler is exclusively trained on reanalysis data and does not undergo a fine-tuning step on operational analyses. Its performance during inference is satisfactory; however, future work could investigate whether fine-tuning on operational analyses or even on forecast data could further improve the hourly forecasts. 

Furthermore, we acknowledge a structural limitation of the current design: the 6-hourly temporal resolution of the forecaster does not allow us to fully exploit the high-resolution information of our 1\,\unit{km} grid, because of a mismatch in the characteristic spatio-temporal scales of the processes involved. Most processes for which kilometre-scale information is useful take place and remain predictable only at timescales much shorter than 6\,\unit{h}. Because the forecasting model bypasses these timescales entirely, it has little incentive to incorporate that information in its latent space. We speculate that this might be detrimental to the forecaster's performance as well as, indirectly, to that of the temporal downscaler. Consequently, other designs are being explored, such as training a single forecasting model predicting multiple hourly steps jointly in each autoregressive step. 

Another notable limitation is Varda-single's sub-optimal performance over complex terrain. In the regional domain, the model ingests data on a 1\,\unit{km} grid, and the encoder is required to map fine-scale information into the latent space without losing relevant details. During decoding, each 1\,\unit{km} output cell is reconstructed from its three nearest latent nodes. Consequently, multiple output cells share the same latent nodes, requiring a strong expressive capacity in the latent representation. In regions with complex topography, neighbouring cells can have very different weather conditions despite relying on the same latent nodes, making the decoding even harder. We therefore hypothesise that the current latent capacity may be insufficient to represent the fine-scale variability of such regions. Future work should investigate the capacity of the latent representation in terms of spatial coverage and the number of channels. A promising research direction could involve dynamically increasing the density of the latent nodes in more demanding regions, therefore helping the model by introducing an inductive bias towards complex terrain.

Finally, we plan to address the over-smoothing behaviour of deterministic training by adopting spatially aware loss formulations, such as the one proposed by \citet{subich2025}. This limitation will also be addressed by training probabilistic models, which is a separate active line of work.

These developments will further build on the framework presented here, Varda-single-1.0, which is already available for broader use. Model weights and configuration files are openly available on Hugging Face, together with setup instructions and examples for producing seamless hourly forecasts over Switzerland at 1\,\unit{km} resolution for lead times of several days, on consumer-grade hardware. 

Varda-single is now running quasi-operationally at MeteoSwiss with a 3\,\unit{h} cadence, with forecasts soon to be made available to the public.

\codeavailability{The model weights as well as training and inference configurations for the Varda-single forecasting and temporal interpolation components are available at \url{https://huggingface.co/MeteoSwiss/Varda-single-1.0}, which also includes examples of how to run the models with near-real-time data. The Anemoi framework on which Varda-single is built is fully open source; see \url{https://github.com/ecmwf/anemoi}. The evaluation code used in this study is available in the open-source \texttt{evalml} package : \url{https://github.com/MeteoSwiss/evalml}.} 






\newpage
\appendix

\clearpage

\section{Machine Learning Components} \label{app:ml-components}

Varda-single relies on the Anemoi framework for model architecture and training. Anemoi builds on PyTorch \citep{ansel_pytorch_2024}, PyTorch Lightning \citep{falcon_pytorch_2019}, and PyTorch Geometric \citep{fey_pyg_2025}.

We briefly describe the graph-based layers in Varda-single using the message passing (MP) framework \citep{gilmer2017icml}, where each node aggregates learnable vector-valued messages computed for each node in its neighbourhood (typically the 1-hop neighbourhood), before updating its own representation.
The MP layers in Varda-single are Graph Transformers, a family of anisotropic functions that weights messages with the attention mechanism of \cite{vaswani2017anips} before aggregating them. The specific architecture implemented in Anemoi is adapted from \cite{shi2021}. The message function producing a message $m_{i,j}$ for each edge $(i,j)$ is defined as:
\begin{equation}\label{eq:graph-transformer-msg}
\begin{aligned}
    & m_{i,j}  = \alpha_{i,j}\left( W_v h_j + W_e e_{ij}\right) \\
    & \alpha_{i,j} = \mathrm{softmax}_{j \in \mathcal{N}(i)} \left(\frac{(W_{q} h_i)^T (W_{k}^{\text{n}} h_j+W_e e_{ij})}{\sqrt{d}}\right)
\end{aligned}
\end{equation}
Here, $h_{i}$ denotes the hidden representation of node $i$, $\mathcal{N}(i)$ denotes its neighbourhood, and $e_{ij}$ represents edge features. $W_q$, $W_k$ and $W_v$ are the linear projections that transform node embeddings into queries, keys and values respectively. $W_e$ is the linear projection for edge features, and $W_s$ is the linear projection for the self-connection term. Each node then aggregates the messages with a simple sum and updates its representation with a feed-forward neural network (FFNN) with skip connections and layer norms.
As in the standard Transformer architecture and the model of \cite{velickovic2018}, the attention mechanism can have multiple attention heads. In simple terms, each node aggregates information from its neighbours as a learnable weighted average, such that the weights of all incoming messages sum to $1$. Overall, the layer can be summarised as:

\begin{equation}\label{eq:graph-transformer-aggr}
\begin{aligned}
      h_i^{l+1} = \mathrm{FFNN}_{\theta}^l \left( h_i^l,\ W_s h_i^l + \sum_{j \in \mathcal{N}(i)} \alpha_{i,j}\left( W_v h_j^l + W_e e_{ij}^l \right) \right)
\end{aligned}
\end{equation}
which is used in the encoder, processor and decoder blocks of the overall model.

\section{Graph Structure}
\label{app:graph-structure}

Both the forecaster and the temporal downscaler operate on a graph composed of two node sets and
three edge sets. The node sets consist of data nodes $V_D$, which correspond to the input grid, and hidden nodes $V_H$, which are used
to represent the latent state. The edge sets connect data-to-hidden, hidden-to-hidden, and hidden-to-data nodes, and
define the three graphs the model operates on:
$\mathcal{G}_{\mathrm{enc}} = (V_D \cup V_H,\, E_{\mathrm{enc}})$,
$\mathcal{G}_{\mathrm{proc}} = (V_H,\, E_{\mathrm{proc}})$ and
$\mathcal{G}_{\mathrm{dec}} = (V_H \cup V_D,\, E_{\mathrm{dec}})$.
The data-to-hidden and hidden-to-data sets each form bipartite sub-graphs between the data and hidden nodes. All edges carry two geometric attributes: normalised edge length and normalised edge direction. The models process the edge sets sequentially in distinct
message-passing stages.

\begin{figure*}[tbp]
    \centering
    \includegraphics[width=\linewidth]{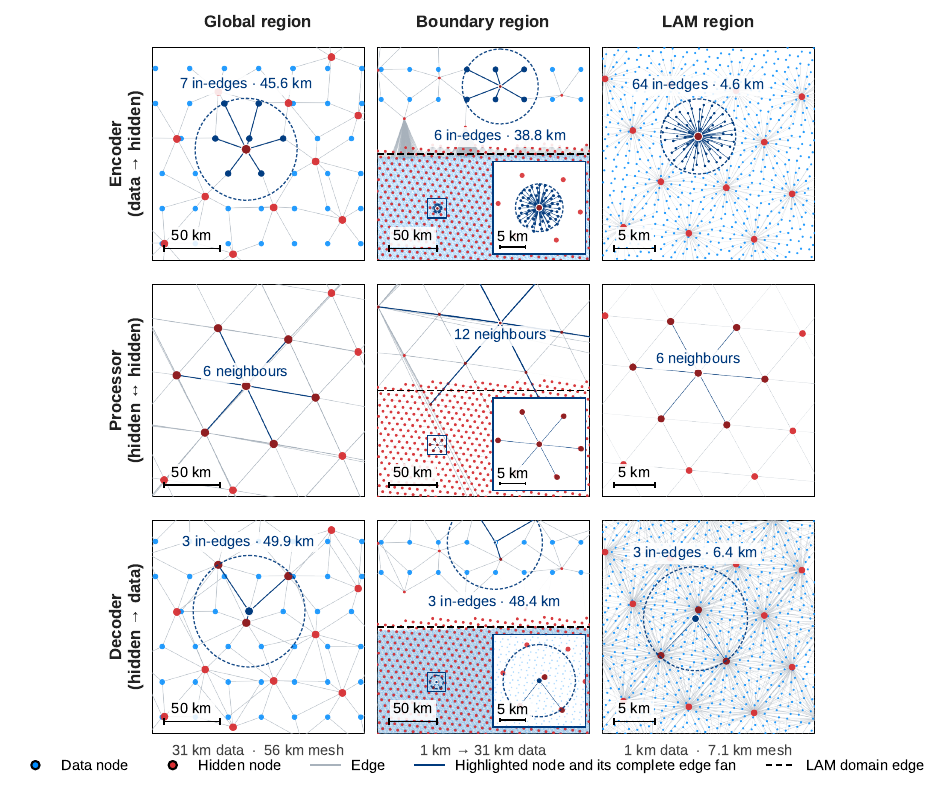}
    
    \caption{Graph connectivity shown across three representative regions (left to
    right in each row: global ERA5 region, boundary between global and LAM, LAM-only
    region). Blue dots are data nodes, red dots are hidden nodes, and grey lines are
    edges.}
    \label{fig:connectivity}
\end{figure*}

A visual representation can be seen in \autoref{fig:connectivity}.

\textbf{Hidden mesh construction.}

The hidden nodes $V_H$ are placed on a triangular icosahedral mesh, obtained by
recursively subdividing the faces of a base icosahedron. Each refinement level
quadruples the number of faces and roughly halves the characteristic edge length, so
the refinement level directly sets the spatial resolution of the latent
representation. During global pre-training the mesh is refined uniformly to a single
global level, yielding a quasi-homogeneous node distribution over the sphere. For
stretched-grid training and fine-tuning the mesh is refined to a coarse level over the
global domain and to a finer level within the limited-area domain, with a small margin
around the regional boundary so that the density transition does not introduce
discontinuities in connectivity. The resulting mesh is dense over the region of
interest and coarse elsewhere, matching the multi-resolution nature of the training
data while keeping the global node count tractable. The global and regional refinement
levels used at each training stage are reported in \autoref{sec:graph}.

\textbf{Encoder edges $E_{\mathrm{enc}}$.}

The encoder edges connect each hidden node to a local neighbourhood of data nodes. We
consider two connectivity schemes: k-nearest-neighbours and radial cutoff.  In the \emph{$k$-nearest-neighbours} (KNN) scheme,
each hidden node is connected to its $k$ nearest data nodes; this yields a fixed
in-degree but does not account for the varying node density of a stretched mesh, where
a fixed $k$ over-connects the coarse global region and under-connects the dense
regional domain. In the \emph{radial cutoff} scheme, an edge is created between a data
node and a hidden node whenever their great-circle distance is below a threshold
$d_{\mathrm{cut}} = \alpha\,\bar{d}_H$, where $\bar{d}_H$ is a reference hidden-node
spacing and $\alpha$ is a dimensionless cutoff factor. Since $\bar{d}_H$ scales with
the local mesh resolution, the cutoff radius produces a consistent receptive field
across the resolution transition. We use $\alpha = 0.7$ in all stretched-grid
experiments. 

\textbf{Processor edges $E_{\mathrm{proc}}$.}

The processor edges are constructed as \emph{multi-scale} edges on the icosahedral
mesh. At every refinement level of the mesh hierarchy, each hidden node is connected to
its immediate ($1$-hop) neighbours on that level; the union of these level-wise
neighbourhoods forms $E_{\mathrm{proc}}$. These edges are undirected, so each hidden node has a
single neighbourhood rather than separate incoming and outgoing sets, and that neighbourhood is not
of fixed size: a node belonging to $L$ refinement levels gains roughly six neighbours per level, so
neighbourhood sizes cluster near multiples of six and grow with the number of levels a node takes
part in. In the stretched-grid graph 75\,\% of hidden nodes have six neighbours and 19\,\% have
twelve, while the remaining 6\,\% have up to 54 neighbours. \autoref{fig:connectivity} shows representative
nodes with six and twelve neighbours; the nodes carrying the longest edges have larger
neighbourhoods than any of those shown. Because coarser levels connect nodes that are
far apart on the sphere while finer levels connect nearby nodes, the connectivity spans
a range of length scales, letting the processor exchange information both locally and
over long distances without a deep stack of purely local hops, keeping the effective
graph diameter low while preserving a global receptive field.

\textbf{Decoder edges $E_{\mathrm{dec}}$.}

Each data node is connected to its three nearest hidden nodes, so its output is obtained
by interpolating the latent state of the surrounding mesh nodes. A small fixed number
of neighbours is sufficient because the latent mesh is, by construction, at least as
dense as the data grid over the region of interest, making the decoding a local
interpolation.

\textbf{Edge attribute normalisation.}

The normalised edge length is the great-circle distance between source and target
nodes scaled to $[0,1]$ by the maximum edge length within its edge set (unit-max
normalisation), while the edge direction is standardised to zero mean and unit variance
across the edge set (unit-std normalisation).

\section{Hardware details}

Training of the operational emulator was performed on up to 16 nodes of the Balfrin cluster, part of the Alps supercomputer at the Swiss National Supercomputing Centre (CSCS). Each node is equipped with 4 NVIDIA A100 GPUs, each with 96 GB of memory, for a total of up to 64 GPUs used in training.

The temporal downscaler was trained on 16 nodes with 4 NVIDIA GH200 GPUs each, for a total of 64 GPUs. This training was carried out on the Santis cluster, also part of the Alps supercomputer infrastructure at CSCS.

Most of the development and experimentation phase was instead carried out on the LUMI supercomputer (Large Unified Modern Infrastructure). Each node is equipped with 4 AMD Instinct MI250X GPUs; since each MI250X module comprises two compute dies exposed to the runtime and scheduler as independent devices, each node provides 8 GPU devices, each with 64 GB of memory, for a total of 128 GPUs.

\section{Geographical information}

\begin{figure}[!htbp]
    \centering
    \includegraphics[scale=1]{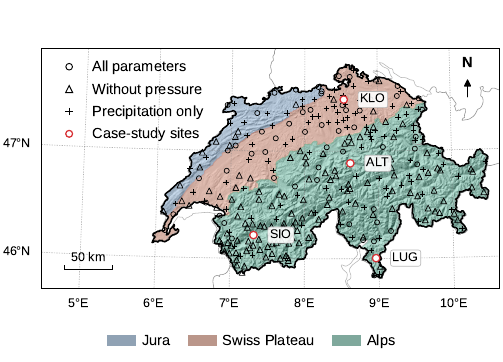}
    \caption{Map of Switzerland showing the Jura Mountains, the Swiss Plateau and the Swiss Alps, together with the surface stations used in the evaluation. There are a total of 187 stations measuring 2\,\unit{m} temperature, 151 stations measuring 10\,\unit{}{m} wind speed and direction, 59 stations measuring mean sea-level pressure and 277 stations measuring total precipitation.}
    \label{fig:geography-stations}
\end{figure}

The geographical regions used for evaluation in this work are the Jura Mountains, the Swiss Plateau, and the Swiss Alps, as shown in \autoref{fig:geography-stations}. The Jura region is characterised by sub-alpine mountain ranges with elevations of up to approximately 1,700\,\unit{m}. The Swiss Plateau is a relatively low-lying, mostly hilly region between the Jura and the Alps, with elevations ranging approximately from 400 to 800\,\unit{m}. The Swiss Alps are characterised by complex, high-elevation terrain, with pronounced valleys and peaks reaching elevations of up to 4,500\,\unit{m}. The figure also shows the surface stations of the SwissMetNet network \citep{swissmetnet2026} used in the verification of the models against measurements. We use measurements of 2\,\unit{m} temperature, 10\,\unit{m} wind speed and direction, mean sea-level pressure and precipitation.

\noappendix       






\authorcontribution{\mbox{}\par\nobreak
\begin{itemize}
    \item Data curation and dataset preparation: FZ, MA, LJ, DL
    \item Funding acquisition: UH, ML, OF
    \item Conceptualization: ML, OF
    \item Model development and training: AP, MC, RR, OM
    \item Software: AP, FZ, MC, CM, JB, LF, HdL, OM, CO, DN, DH, MM, GM, APr, MCh, HH, MSC, JW, GM, HC
    \item Supervision: AP, FZ, CM, JB, CO, DN, OF
    \item Validation and evaluation: AP, FZ, CM, JB, LF, OM, CO, DN, APa, MA, GM
    \item Visualization: AP, MC, CM, RR, JB, LF
    \item Model integration and dissemination: FZ, HdL, DN
    \item Project coordination and administration: FZ, CM, CO, DN, ML, KE, APr, MCh, OF
    \item Writing (original draft preparation): AP, FZ, MC, JB, LF, CO, DN
    \item Writing (review and editing): AP, FZ, MC, CM, RR, JB, LF, CO, DN, APa, MB, LL, OF
\end{itemize}
} 

\competinginterests{The contact author has declared that none of the authors has any competing interests.} 


\begin{acknowledgements}
We gratefully acknowledge the many colleagues whose contributions supported the development of Varda and the work presented here. We particularly thank David Leutwyler, Jean-Marie Bettems, Petra Baumann, Jan Pfirter, Francesco Isotta, Dominik Büeler, Christian Grams, Pirmin Kaufmann, Christian Kanesan and Nina Burgdorfer for their contributions to the development, processing and validation of REA-L-CH1 and related data infrastructure. REA-L-CH1 provides an important foundation for the development and evaluation of Varda. Their contributions include model-development input, dataset preparation, validation against observational datasets, software support and resolution of data and grid issues.

We thank Alberto De Stefani, Thomas Jordi, Adrien Michel, Leonard Knirsch, Matjaz Puh, Simon Adamov, Tobias Wicky, Capucine Lechartre and Matteo Buzzi for contributions to evaluation tooling, data delivery and preparation, scientific analysis, earlier emulator development, and inference and output workflows that supported the broader Varda effort.

We acknowledge Christoph Spirig, Sünje Dallmeier-Tiessen, Xavier Lapillonne, Gabriela Aznar, Lionel Moret, Michele Volpi and Ulrich Hamann for coordination, infrastructure, funding and organisational support that enabled the project.

Varda builds on the Anemoi open-source ecosystem and on extensive collaboration with the wider Anemoi community. We thank all Anemoi contributors, and in particular Magnus Ingstad, Mariana Clare, Simon Lang, Cathal O’Brien, Dieter Van den Bleeken, Michiel van Ginderachter, Baudouin Raoult, Florian Pinault, Rilwan Adewoyin, Helen Theissen, Jesper Dramsch, Jan Polster, Sara Hahner, Sophie Buurman and Aram Farhad for technical advice, software development, datasets, testing and support that contributed to the broader work on which Varda builds.

We acknowledge the Swiss National Supercomputing Centre (CSCS) for providing access to the HPC facilities of the Alps supercomputer. We acknowledge the EuroHPC Joint Undertaking for awarding this project access to the EuroHPC supercomputer LUMI, hosted by CSC (Finland) and the LUMI consortium through a EuroHPC Regular Access call.

We acknowledge the broader collaboration between MeteoSwiss and the Swiss Data Science Center (SDSC), formalized through their framework agreement to strengthen the use of artificial intelligence and machine learning in meteorology and climatology.
\end{acknowledgements}

\financialsupport{This work benefited from financial support provided through the ECMWF Member and Co-operating States Machine Learning Project, which facilitated scientific exchanges and working visits among participating institutions. Michele Cattaneo was supported by the EUMETNET Artificial Intelligence and Machine Learning for Weather, Climate and Environmental Applications (E-AI) Programme through the 2025 Research Cooperation Grant.}







\newpage
\bibliographystyle{copernicus}
\bibliography{references.bib}

\end{document}